\RequirePackage{fix-cm}
\documentclass[smallextended]{svjour3}       
\smartqed  
\usepackage{graphicx}
\usepackage{newtxtext,newtxmath}      

\usepackage{latexsym}
\usepackage{natbib}
\usepackage{url}
\usepackage{enumitem}
\usepackage{multirow}
\usepackage{makecell}
\usepackage{longtable}
\usepackage{booktabs}
\usepackage{caption}
\usepackage{subfig}
\usepackage{float}
\usepackage{hyperref}
\usepackage{tcolorbox}
\newtcolorbox{summarybox}[1]{ 
  colback=white,
  colframe=black,
  coltitle=black,
  colbacktitle=lightgray,
  width=0.99\columnwidth,
  boxrule=0.5pt,
  arc=2mm,
  title=\textbf{#1},
  center
}

\begin{document}


\title{Moderately Mighty: To What Extent Can Internal Software Metrics Predict App Popularity at Launch?}
\titlerunning{Moderately Mighty: App Popularity Prediction with Internal Software Metrics}

\author{Md Nahidul Islam Opu \and
        Fatima Islam Mouri \and
        Rick Kazman \and
        Yuanfang Cai \and
        Shaiful Chowdhury
}

\institute{Md Nahidul Islam Opu \and Fatima Islam Mouri \and Shaiful Chowdhury\at
    SQM Research Lab, University of Manitoba, Winnipeg, Canada\\
    \email{opumni@myumanitoba.ca \hspace{6mm} fimouri.4ws@gmail.com \hspace{6mm}shaiful.chowdhury@umanitoba.ca}
    \and
    Rick Kazman \at
    University of Hawaii, Honolulu, USA\\
    \email{kazman@hawaii.edu}
    \and
    Yuanfang Cai \at
    Drexel University, Philadelphia, USA\\
    \email{yfcai@cs.drexel.edu}
}

\date{Received: date / Accepted: date}

\maketitle

\begin{abstract}
Predicting a mobile app's popularity before its first release can provide developers with a strategic advantage in a competitive marketplace, yet it remains a challenging problem. This study explores the extent to which internal software metrics, measurable from source code before deployment, can predict an app’s popularity (i.e., user ratings and downloads per year) at its inception. For our analysis, we first constructed a rigorously filtered dataset of 446 open-source Java-based Android apps that are available on both F-Droid and Google Play Store. Using app source code from F-Droid, we extracted a wide array of internal metrics, including system-, class-, and method-level code metrics, code smells, and app metadata. Popularity-related information, including reviews and download counts, was collected from the Google Play Store. 

We evaluate different regression and classification models across three feature sets: a minimal \texttt{Size-only} baseline, a domain-informed \texttt{Handpicked} set, and a \texttt{Voting} set derived via different feature selection algorithms. Our results show that, for both app ratings and number of downloads, regression models perform poorly due to skewed rating distributions and a highly scattered range of download counts in our dataset. However, when the problem is reframed as a binary classification task (Popular vs. Unpopular), performance improves significantly---the best model, a Multilayer Perceptron, achieves an F1-score of 0.72. 

We conclude that, although internal code metrics alone are insufficient for accurately predicting an app’s future popularity, they do exhibit meaningful correlations with it. Thus, our findings challenge prior studies that have entirely dismissed internal metrics as valid indicators of software quality. Instead, our results align with research suggesting that internal code metrics can be valuable when evaluated within the appropriate context—specifically, we found them useful for classification tasks. 

\end{abstract}

\section{Introduction}

The rapid evolution of handheld devices, such as smartphones and tablets, along with the development of mobile operating systems, has broadened the scope, complexity, and opportunity in mobile application development. As the user base continues to grow, so does the demand for mobile applications, leading to an increase in the number of vendors~\citep{fu2013people}. Mobile application developers primarily rely on prominent platforms such as Google Play and Apple's App Store to distribute their applications to end users. These platforms provide a centralized marketplace where developers can sell or distribute their products, and users can compare applications, read reviews and ratings, and make informed decisions—essentially serving as "one-stop shops" \citep{zhu_what_2024}. Users contribute to the marketplace by providing reviews and ratings that reflect their level of satisfaction or dissatisfaction with an application based on personal experience. 

When choosing applications, users consider a variety of factors. While features such as being \emph{ad-free} or \emph{free-to-use} are appealing, the most influential indicators of an app's popularity are its user reviews (e.g., ratings) and the number of downloads~\citep{alhejaili_study_2022,liptrot_why_2024}. These metrics play a critical role not only in shaping user perception but also in determining the commercial success of developers who rely on app sales for income.

Given the strong connection between a developer's success and an app's popularity, we argue that developers would greatly benefit from the ability to predict an app’s popularity prior to its release. Such foresight is especially valuable before launching the initial version, as users often form lasting impressions based on their first experience~\citep{digirolamo_first_1997,human_accurate_2013,miller_are_2004}. Additionally, the lower number of early downloads and user ratings can significantly influence app store ranking algorithms~\citep{karagkiozidou_app_2019}, affecting the app’s visibility and making it more difficult to attract users later—even if the app is improved. Unfortunately, existing relevant work required history data while analyzing and predicting app popularity, making such analysis impractical for popularity prediction at an app's inception \citep{catolino_does_2018}. 

Previous research~\citep{ali2017same} observed that Android users participate more in rating apps than iOS users, making the Android app stores more suitable for our study. Leveraging data from these stores, this paper investigates the extent to which app popularity can be predicted using only internal software metrics available at the early stages of development---specifically, at an app's inception. Prior research has demonstrated that internal metrics can reliably estimate external software quality attributes, such as bug-proneness, change-proneness, and the ease of extending functionality. Given that these external factors are strongly linked to app popularity, we pose the following question: \emph{can internal software metrics be used to predict key indicators of popularity, such as user ratings and download counts?}

While app ratings and download counts are readily available on the Google Play Store, measuring internal software metrics requires access to the app’s source code. Unfortunately, the Play Store does not provide source code, making it infeasible to extract these metrics directly from that platform. To overcome this limitation, we utilized the F-Droid repository\footnote{https://f-droid.org/en/ last accessed: Apr 14, 2025}---a collection of open-source Android applications. By mapping F-Droid apps to their counterparts on the Play Store, we were able to collect both independent (internal software metrics) and dependent variables (ratings and downloads), enabling us to build predictive machine learning models. For internal metrics, we focused on popular code metrics at three granularity levels: system level (e.g., DL score), class level (e.g., C\&K metrics), and method level (e.g., McCabe). Additionally, we extracted commonly used code smells (e.g., excessive dependency) to capture aspects of code and design quality. Since app popularity may also be influenced by factors such as feature count, app genre, and the presence of advertisements, we incorporated these attributes into our analysis as well. 

We aim to investigate the potential of internal code metrics in predicting app popularity, using two widely adopted indicators: user rating and download count. Accordingly, our study is guided by the following two research questions.

\textbf{RQ1: \textit{To what extent can internal software metrics predict an app’s user rating?}}

To address this question, we first applied different regression algorithms to predict an app’s average user rating based on its internal software metrics. Our results indicate that internal software metrics measured from the initial app version perform very poorly in predicting an app's future rating. However, to observe if internal metrics have \textit{any} impact on app popularity, we reformulated the problem as a binary classification task---labeling apps as \emph{Popular} or \emph{Unpopular} based on a predefined rating threshold. In this setting, the results are more promising with an F1-score of 0.72.  This suggests that an app's internal code and design quality are important factors in eventual app popularity, but not the only important factor. This is not surprising given that app popularity may also depend on aspects such as the types of features offered and how pleasing its UI is, which are unlikely to be correlated with any internal software metrics.

\textbf{RQ2: \textit{To what extent can internal software metrics predict an app’s number of downloads?}}

Following a similar approach to RQ1, we first conducted regression analysis on the yearly download count (i.e., downloads per year) rather than the total number of downloads, as the latter can be heavily influenced by an app's age. Consistent with our findings for user ratings, we observed that predicting the exact number of yearly downloads is extremely challenging.  However, when we reframed the task as a binary classification problem---i.e., distinguishing between \emph{Popular} and \emph{Unpopular} apps based on a predefined number of downloads---the results are more encouraging with an F1-score of 0.69. 


In summary, while internal metrics alone are insufficient to accurately predict an app’s future popularity, they do offer meaningful predictive insights (hence the title, \emph{moderately mighty}). This finding challenges earlier studies~\citep{el_emam_confounding_2001,gil_correlation_2017} that dismissed the utility of such metrics entirely. In line with another stream of research~\citep{chowdhury_revisiting_2022,landman_empirical_2014}, we find that \emph{context is king} when evaluating internal metrics---observing a negative result in one context (e.g., regression, in our case) does not preclude encouraging results in others (e.g., classification). Our results, therefore, reaffirm the importance---albeit in a limited capacity---of monitoring a variety of code-quality indicators. By detecting early warning signs, developers can proactively improve their apps, potentially saving time, resources, and reputation. Similarly, app stores could leverage internal metrics to enhance ranking algorithms and allocate support more effectively.


To enable replication and extension, we share our data publicly.\footnote{\url{https://github.com/SQMLab/AppPopularityPrediction}}

The remainder of this paper is structured as follows: related work is discussed in Section \ref{sec:related_works}; methodology is described in Section \ref{sec:methodology}; and approach, analysis, and results for the research questions are presented in Section \ref{sec:results}. We discuss the implications, threats and future works in Section \ref{sec:discussion}. Finally, we conclude the paper in Section \ref{sec:conclusion}.

\section{Related Work and Motivation}
\label{sec:related_works}
In this section, we review previous research on App Store analysis. We establish why building internal metrics-based models for predicting app popularity is still an open research problem. We then discuss related work showing the potential of code metrics,  code smells, and app metadata in building such models.   


\subsection{App Popularity Analysis}

Ali et al.~\citep{ali2017same} analyzed apps available on both the Play Store and App Store, finding that Android users are more active in rating apps than iOS users. Their study also revealed that majority of the most popular apps are available across both platforms. In a separate analysis of the App Store, Harman et al.~\citep{harman2012app} reported no significant correlation between an app’s price and its rating or popularity. Nayebi et al.~\citep{nayebi2018app} highlighted the value of incorporating user feedback from external sources, such as Twitter, in addition to App Store reviews. Ickin et al.~\citep{ickin2017users} observed that while users often rely on reviews when choosing to install apps, they tend to uninstall them due to issues like frequent crashes, intrusive advertisements, and lack of ongoing maintenance, such as feature updates. The problem of crashes and lack of ongoing maintenance issues were also observed by Khalid et al.~\citep{khalid_what_2015} while analyzing common complaints in app reviews. 

Corral et al.~\citep{corral_better_2015} examined the influence of C\&K metrics on mobile app popularity, finding only a weak correlation. However, their study relied on displayed star ratings, an inaccurate approach as ratings can vary by region, app version, and device type~\citep{google_google_2024,apple_ratings_2024}. Similarly, Gezici et al.~\citep{gezici2019internal} reported no significant association between internal code quality and external app quality. In contrast, Cruz et al.~\citep{cruz2019attention} argued that thorough testing can enhance code quality, which may in turn improve user ratings. The relationship between code quality and app popularity also appears to be app category–dependent, as observed by Sorbo et al.~\citep{di2021investigating}. Noei et al.~\citep{noei_study_2017} found that both device-specific attributes (e.g., memory capacity, display size, battery life) and app-specific characteristics (e.g., UI complexity, code size) can influence user-perceived quality in Android apps. Similarly, Tian et al.~\citep{tian2015characteristics} observed that, compared to low-rated apps, high-rated apps tend to be larger in size, exhibit greater complexity, and benefit from more extensive marketing efforts.  
Interestingly, Kuttal et al.~\citep{kuttal2020tug} highlighted a disconnect between developer and user perspectives: they found a weak correlation between an app’s popularity on GitHub and its ratings in the App Store.

Gezici et al.~\citep{gezici2019neural} showed that user ratings can be accurately predicted from user reviews using neural sentiment analysis. Rahman et al.~\citep{rahman2017predicting} found that static code metrics can be employed to build models for predicting privacy and security issues in Android apps. In particular, using a radial-based support vector machine algorithm, they achieved a precision of 0.83. However, this high precision came at the expense of lower recall. Similar results were reported by Alenezi et al.~\citep{alenezi2018empirical}, who also used code metrics to predict risk scores. 

Grano et al.~\citep{grano_android_2017} extracted various code metrics and code smells from decompiled APKs to support research on software evolution and quality improvement. Their methodology involved collecting app versions and user reviews from F-Droid and Google Play, and assessing code smells and quality indicators using automated tools. Catolino~\citep{catolino_does_2018} built upon the dataset of Grano et al.~\citep{grano_android_2017} to examine whether code quality could be used to predict app ratings. Although the objectives of Catolino’s study align with ours, it is a brief (2-page) work without much in-depth analysis and suffers from two significant methodological issues, impacting the reliability of its findings.

\begin{itemize}
    \item The study relied on the dataset of Grano et al.~\citep{grano_android_2017}, which derived code quality metrics from decompiled APKs. However, decompiled code is not equivalent to the original source code, as compilers introduce various optimizations during the APK build process~\citep{you_comparative_2021,zeng2019studying}. Consequently, critical code metrics—such as code readability—may differ significantly between the original and decompiled versions. More concerningly, decompiled code often differs from the original source code both syntactically and semantically~\citep{harrand_strengths_2019}, thereby compromising the validity of any analysis based on metrics derived from it.

    \item To evaluate model performance, Catolino adopted a 10-fold cross-validation approach. This method is fundamentally flawed in scenarios involving multiple versions of the same entity, which is true for the used dataset. In such scenarios, 10-fold cross-validation approach introduces data leakage: some versions may end up in the training set while others appear in the test set. This leakage leads to overestimated model performance. Pascarella et al.~\citep{pascarella_performance_2020} and Chowdhury et al.~\citep{chowdhury_method-level_2024} have examined this issue in detail. Both studies revisited highly accurate method-level bug prediction models and demonstrated that their performance dropped significantly when evaluated using more realistic approaches that prevent overlap between training and test data.
\end{itemize}

Based on our previous discussion, we identified a subset of research that explored correlations between app attributes and app popularity. However, these findings were often inconsistent, with both positive and negative correlations reported. Moreover, only a limited number of studies attempted to develop predictive models based on app attributes, and those efforts were undermined by methodological flaws that compromised their reliability. \emph{Motivated by these gaps, we conduct a new, rigorous study to assess how accurately app popularity can be predicted using only internal software metrics. Importantly, we focus exclusively on the first release of each app, ensuring that our model does not rely on historical data, making it particularly valuable for developers seeking to build a strong reputation from the outset.}

\subsection{Code Metrics}
Code metrics—quantitative measures derived from analyzing source code—have long served as a foundation for assessing software quality and maintainability~\citep{mccabe_complexity_1976,Alves:2010,pascarella_performance_2020,chowdhury_method-level_2024,mashhadi_empirical_2024,ferenc_deep_2020,alsolai_systematic_2020}. A key advantage of code metrics is their availability throughout the software development life cycle, which allows for the construction of predictive maintenance models even in the absence of historical data, effectively addressing the cold start problem~\citep{chowdhury2025good,pascarella_performance_2020}. Researchers have developed and studied code metrics at different granularities while analyzing their effectiveness in understanding software maintenance. 

\textbf{\emph{System-level metrics}}. System-level metrics play a pivotal role in evaluating the long-term maintainability of software systems~\citep{liu_prevalence_2024}. These metrics provide valuable insights into the overall system architecture and its degradation patterns over time~\citep{mo2016decoupling,rachow_architecture_2022,liu_prevalence_2024}. Architectural issues at the system level not only hinder quality but also escalate the cost and complexity of future enhancements. 
To address these challenges, researchers have proposed various system-level metrics. For instance, Sethi et al.~\citep{sethi2009retrospect} introduced the Independence Level metric, while Mo et al.~\citep{mo2016decoupling} developed the Decoupling Level metric. Both aim to assess how effectively a software system can be modularized into independent components. Similarly, MacCormack et al.~\citep{maccormack2006exploring} proposed the Propagation Cost metric, which measures the degree of coupling among source files within a system. Empirical studies have shown that systems with low decoupling scores tend to be more bug-prone and harder to extend with new functionalities~\citep{mo2016decoupling}. 
Further advancing this area, Cai et al.~\citep{cai_software_2023} discussed novel architecture-level maintainability metrics and identified architectural design anti-patterns that serve as early warnings of potential structural issues. These anti-patterns assist developers in detecting and resolving architectural flaws before they evolve into critical maintenance challenges.

\textbf{\emph{Class-level and method-level metrics}}. Some of the most widely adopted class-level metrics are the Chidamber and Kemerer (C\&K) metrics~\citep{Chidamber:1994}, which include measures such as depth of inheritance, coupling between objects, and number of children. Basili et al.~\citep{basili1996validation} demonstrated that C\&K metrics are effective predictors of class fault-proneness during the early stages of software development. Similarly, Li et al.~\citep{li1993object} found these metrics useful for estimating maintenance effort. A broader range of studies supporting the predictive power of C\&K metrics for maintainability was summarized by Iftikhar et al.~\citep{iftikhar2024tertiary}. However, not all findings were consistently positive---some studies, such as Gil and Lalouche~\citep{gil_correlation_2017}, reported limited usefulness of class-level metrics in estimating future maintenance efforts.

In recent years, there has been a growing focus on more fine-grained, method-level code metrics~\citep{chowdhury_method-level_2024,chowdhury2025good,pascarella_performance_2020,chowdhury_revisiting_2022}, largely driven by developers’ interest in insights at that level of granularity. While Pascarella et al.~\citep{pascarella_performance_2020} reported that method-level metrics performed poorly in method-level bug prediction, Chowdhury et al.~\citep{chowdhury_method-level_2024,chowdhury2025good} found these metrics to be effective indicators of method-level bug-proneness and change-proneness. Additional studies also supported these findings~\citep{landman_empirical_2014,Giger:2012,Mo:2022}, reinforcing the value of method-level metrics for software quality assessment.

\emph{Encouraged by this body of work, we investigate whether code metrics across these three levels of granularity---system, class, and method---can be used to predict app popularity, given their demonstrated success in predicting other software quality and maintainability attributes.}

\subsection{Code Smells}
Code smells are sub-optimal design choices in source code that violate established design principles, reducing code quality and maintainability. While their effects may not be immediately visible, over time, they make the codebase harder to understand, modify, and maintain \citep{zhang_code_2024}. The software engineering community widely recognizes the negative implications of code smells, which often result in higher maintenance efforts \citep{yamashita_exploring_2013,abbes_empirical_2011}. Given their significant long-term impact, code smell detection has become one of the most extensively studied topics in the field of software quality and maintenance~\citep{zhang_code_2024}. As such, a stream of research focused completely on developing code smell detection techniques and tools~\citep{moha_decor_2010,sharma_designite_2016,palomba_lightweight_2017}. While studies exist that found minimal or no impact of code smells, the majority of studies agree that code smells adversely affect software maintenance~\citep{khomh2009exploratory,spadini_relation_2018,zhang2017empirical,palomba_diffuseness_2018,palomba2017toward,Khomh:2012}. For example, Khomh et al.~\citep{khomh2009exploratory} found that classes with code smells tend to be more change-prone than those without any smells. While numerous studies showed the negative impact of source code smells on change- and bug-proneness, Spadini et al.~\citep{spadini_relation_2018} found that code smells negatively impact the change- and bug-proneness of test code as well. 

\emph{Recognizing the already established quality and maintenance impact of code smells, we include different types of code smells in our study while predicting app popularity.}

\subsection{Other Metadata}

In addition to code-level characteristics, prior research has identified several app-level metadata attributes that significantly influence app popularity. These include genre, advertising policy, permissions requested, and the richness of feature sets~\citep{tian2015characteristics,gui_truth_2015,wang_smartpi_2020}. These factors reflect design, usability, and monetization trade-offs that are often perceptible to users even before installation, thereby shaping their preferences and satisfaction levels.

App category or genre is a high-level descriptor of an app’s intended functionality and target audience. Several studies have examined the distributional effect of genre on user ratings, downloads, and ad strategies. Tian et al.~\citep{tian2015characteristics} found that the genre or category of an app is the fourth important factor that impacts the app's rating. Similarly, Mahmood et al.~\citep{mahmood_identifying_2020} observed that genre is one of the most influential variables and has a high impact on the ratings of apps. These findings collectively indicate that app popularity cannot be decoupled from the genre context in which it operates.

Monetization strategies, particularly the inclusion of in-app advertisements, have also been shown to affect user reception. Gui et al.~\citep{gui_truth_2015} conducted an empirical study demonstrating that in-app advertisements significantly increase CPU, energy, and network resource consumption, leading to user complaints and lower ratings. He et al.~\citep{he_investigation_2018} further quantified this effect, showing that apps embedding more than six ad libraries are substantially more likely to receive negative reviews. These findings underscore the tension between revenue generation and user experience, suggesting that an ad-free policy—or a minimally invasive one—can be a distinguishing factor for app popularity.

Permissions requested by an app often raise user privacy concerns, especially when they appear excessive or unrelated to the app’s core functionality. Wang et al.~\citep{wang_smartpi_2020} leveraged user reviews to investigate how users rationalize permission requests, finding that unexplained or intrusive permissions negatively affect user trust and ratings. Jisha et al.~\citep{jisha_mobile_2018} incorporated permission-based risk assessments into app recommendation systems, arguing that lower-risk apps are generally perceived more favorably.

Feature richness, often inferred from app descriptions or inferred capabilities, has also been explored as a determinant of user appeal. Sarro et al.~\citep{sarro_customer_2018} and Liang et al.~\citep{liang_mobile_2017} showed that the number and diversity of claimed features are highly predictive of user ratings. Using NLP and topic modeling, these studies demonstrate that users are more likely to positively rate apps that advertise functional depth and variety, provided these features align with user needs and are well-integrated.

\emph{Motivated by this body of evidence, we systematically captured genre, advertisement policies, requested permissions, and feature richness as part of our metadata collection. These attributes offer complementary explanatory power beyond code-level metrics, enabling a more holistic prediction of app popularity.}

\section{Methodology}
\label{sec:methodology}
\begin{figure}[ht]
    \centering
    \includegraphics[width=\linewidth]{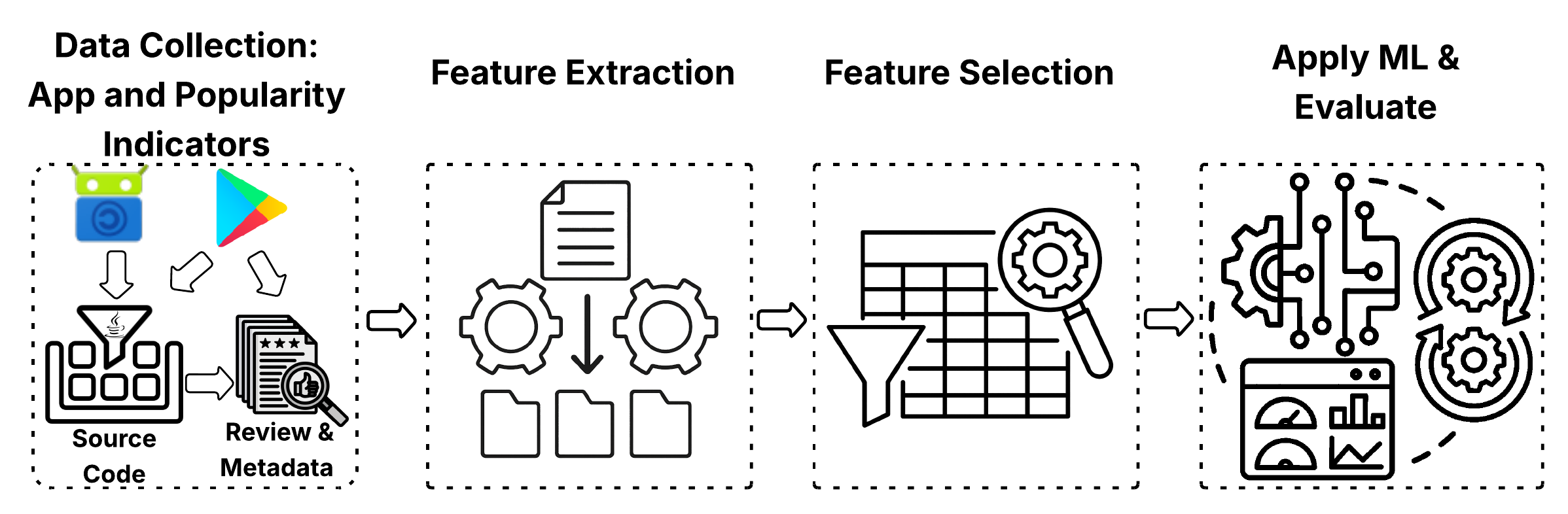}
    \caption{Overview of methodology that involves app collection (with source code and ratings), feature extraction, feature selection, and building predictive models.}
    \label{fig:methodology}
\end{figure}

In this section, we describe our process for app collection, the methodology used for extracting and selecting source code features, the approach for gathering popularity indicators, and the machine learning algorithms employed to build the prediction models. To maintain clarity and readability, methods specific to individual research questions are presented separately in their respective sections (Section~\ref{sec:results}). Figure \ref{fig:methodology} shows the high-level view of our methodology.

\subsection{App Collection}

As we focus exclusively on Android apps, we selected the Google Play Store for collecting popularity metrics---specifically, ratings and number of downloads. Unfortunately, the Play Store does not provide access to source code. Inspired by earlier works in this area~\citep{grano_android_2017,zeng2019studying,coppola2019characterizing}, we leveraged the F-Droid repository\footnote{\url{https://f-droid.org/en/} last accessed: Apr 18, 2025} to collect the source code of the apps. Since our aim is to predict an app's popularity at its inception, we required the earliest available version of each app. F-Droid maintains two repositories: the main repository, which contains the latest three versions of each app, and an archive that stores the remaining versions. We therefore collected information—including package name, source code, GitHub repository URL, and so on—from the oldest available version in the archive. This oldest available version may not represent the application's first version, but it is the closest to the first version that could be reliably collected. Corresponding information, including popularity indicators, genre, advertisement-related information, release date, and other metadata, was retrieved from Google Play. To map apps between F-Droid and Google Play, we used the package name as a common identifier, similar to~\citep{grano_android_2017}. The package name, also known as the application ID, is unique for every application\footnote{\url{https://developer.android.com/build/configure-app-module}}. We further validated this through a manual inspection of 30 applications that share the same package name on both F-Droid and Google Play. For each application, we systematically compared the app icon, description, and promotional images across the two platforms. Although some discrepancies were observed, particularly in cases where not all elements were identical, the similarities were sufficient to unambiguously confirm that the applications are indeed the same. These differences are primarily attributable to the more frequent updates of visual and textual metadata on the Google Play Store. If an app was not available on both platforms, it was excluded from our dataset. After this step, we obtained a collection of 1,347 unique apps.

The apps were developed using various programming languages, such as Java, Kotlin, C\#, and C++. Unfortunately, the tools we used to extract software metrics—e.g., class-level C\&K metrics--- support only Java. Also, it is not recommended to mix multiple programming languages in code metrics-based studies, as code metrics distributions can differ significantly across programming languages~\citep{Zhang:2013,chowdhury_revisiting_2022}. Consequently, we excluded apps developed in non-Java languages. F-Droid provides a link to the source code repository for each application, with these repositories hosted across various platforms—predominantly GitHub, but also including GitLab, Codeberg, BriarProject, and others. Each repository typically features a user interface element that displays the programming languages used in the project along with their corresponding percentage distributions. To systematically extract this information, we developed a web crawler that navigates to each repository and parses the language composition data. From the collected dataset, we applied a filtering criterion: only those projects in which Java accounts for at least 50\% of the codebase were retained for further analysis. This ensured that our code metric measurements reflected the dominant portion of each app. After this filtering step, 553 apps were retained for analysis. 

\subsection{Popularity Indicators}
\label{sec:popularity_indicators}
User-provided star ratings and download counts are key indicators of app popularity~\citep{alhejaili_study_2022,liptrot_why_2024}. However, the rating displayed on the Play Store is not the actual rating. The displayed rating is not the average of all ratings and also varies by region, device type, and app version~\citep{google_google_2024}. In fact, Ruiz et al.~\citep{ruiz2015examining} reported that app store ratings do not consistently reflect changes in user satisfaction. As such, we decided to calculate the average rating instead of choosing the displayed rating. To do this, we collected all the user reviews for an app. A user review contains data related to star rating, user's comment, developer's reply, corresponding app version, etc. 

\begin{figure}[ht]
  \centering
  \subfloat[CDF of size (LOC).\label{fig:loc_cdf}]{
    \includegraphics[width=0.75\textwidth]{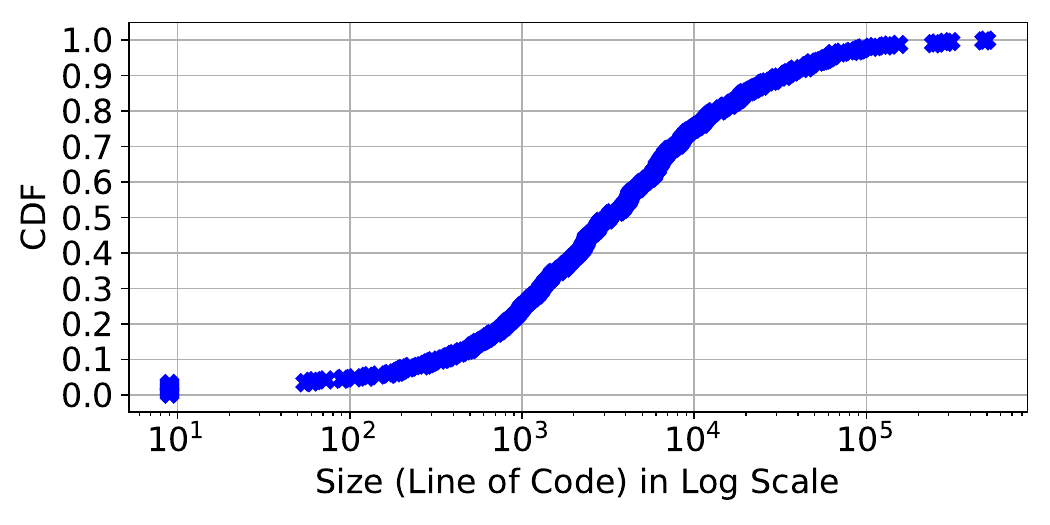}
  }
  \hfill
  \subfloat[CDF of number of Java files.\label{fig:java_files_cdf}]{
    \includegraphics[width=0.75\textwidth]{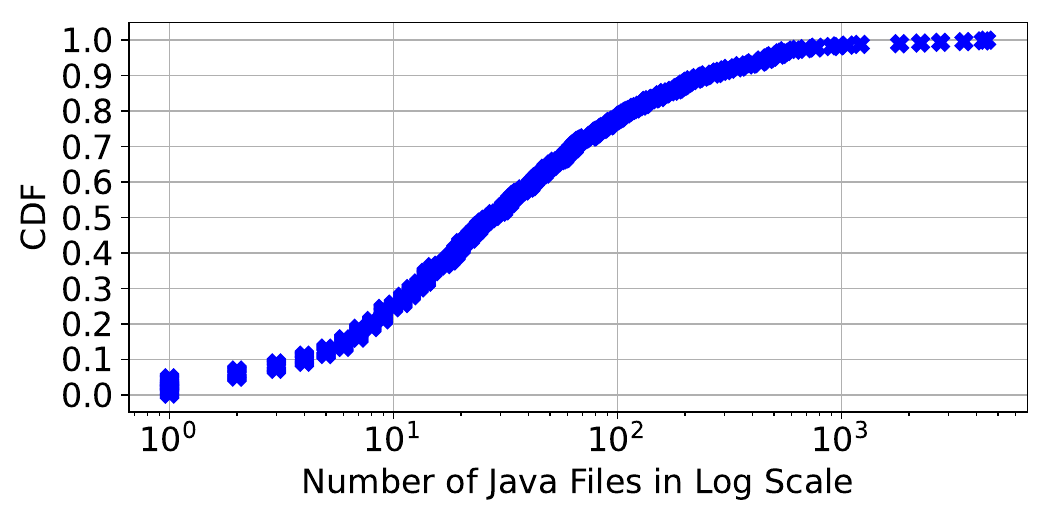}
  }
  \caption{Cumulative distribution functions for (a) the size (LOC) of the applications in log scale and (b) the number of Java files per application in log scale.}
  \label{fig:dataset_overview}
\end{figure}

The use of download count as a standalone metric for assessing application popularity is inherently problematic due to its temporal sensitivity. Specifically, consider two applications, each registering 1,000 downloads—if one was released a year ago while the other has been available for a decade, equating their popularity would be misleading. This disparity arises because download-based popularity measures are intrinsically dependent on the age of the application. To empirically substantiate this hypothesis, we compute Kendall’s tau correlation coefficient between application age and download count. The analysis yields a correlation coefficient of 0.31, accompanied by a p-value below the conventional significance threshold of 0.05. These results provide statistically significant evidence that download count is positively correlated with application age. For this reason, we collected the apps' release data and calculated each app's age. This way, we normalized the age factor and calculated \emph{DownloadsPerYear}. We also discarded apps that are less than 1 year old to enable robust analysis. After this step, we were left with 503 apps. To provide an overview of the dataset, Figure~\ref{fig:dataset_overview} presents the cumulative distribution functions (CDFs) for the size of the applications, measured in lines of code (LOC), and the number of Java files per application.


\subsection{Feature Extraction}
We refer to the analysis of source code and the collection of various quality indicators as feature extraction. To gather features that represent the quality of different aspects of an app, we consider three types of features: code metrics, code smells, and metadata. 
We deliberately exclude platform dependency metrics, device attributes, and performance metrics. Collecting device attributes and performance metrics would necessitate installing and running apps on real devices to gather data, limiting our analysis to a small number of apps and devices. Regarding platform dependency metrics, Syer et al. \citep{syer_studying_2015} demonstrated that defect proneness is associated with platform dependency, specifically the extent to which code relies on Android APIs, but we instead focus on code metrics and design issues that indicate defect proneness. Additionally, managing the ever-evolving Android APIs for all of our collected apps is not feasible. 


\textbf{\emph{System-Level Metrics.}}
Architectural erosion—stemming from poor modularity and unmanaged dependencies—can significantly increase maintenance overhead and degrade software quality over time. To systematically quantify architectural modularity and assess design-time maintainability risks, we employ the DV8 tool~\citep{cai_dv8_2019}, which encapsulates decades of research in software design analysis and education~\citep{cai_software_2023}.

DV8 is a static analysis framework developed to evaluate software architecture through a suite of metrics grounded in design rule theory. It provides actionable insights by identifying architecture anti-patterns and computing maintainability metrics that span structural, evolutionary, and modularity dimensions. The tool supports architecture-level assessment at scale and across programming languages, making it particularly suited for analyzing large software systems in industrial or open-source contexts.

In our study, DV8 was used to extract a total of 17 features that reflect the architectural health of the target software. These include nine quantitative metrics that capture key maintainability dimensions—such as coupling, change impact, and component autonomy—as well as eight anti-pattern indicators that denote architectural flaws observed in the system. Table~\ref{tab:dv8_features} presents an overview of these 17 features, grouped into two categories: architecture maintainability metrics and anti-pattern counts.

\begin{table}[ht]
\centering
\caption{Seventeen System-Level Metrics Extracted Using the DV8 Tool.}
\label{tab:dv8_features}
    \begin{tabular}{p{4.5cm}p{6.5cm}}
        \toprule
           \textbf{Metric} & \textbf{Description} \\
        \midrule
            \multicolumn{2}{l}{\textit{Architecture Maintainability Metrics}} \\
        \midrule
             \textbf{Decoupling Level (DL)} & Degree to which modules are decoupled, reflecting modularity and independence. \\
            \textbf{DL Exclude Isolated Items} & Decoupling level computed while excluding isolated modules. \\
            \textbf{Number of Isolated Items DL Score} & Penalty score reflecting the impact of isolated modules on decoupling level. \\
            \textbf{Number of Files} & Total number of source files analyzed in the system. \\
            \textbf{Propagation Cost (PC)} & Ratio of modules affected by a change, based on dependency propagation analysis. \\
            \textbf{PC Exclude Isolated Items} & Propagation cost excluding isolated modules from the analysis. \\
            \textbf{Number of Isolated Items Score PC} & Weighted score indicating the effect of isolated modules on propagation cost. \\
            \textbf{Independence Level (IL)} & Average structural independence level of all modules in the system. \\
            \textbf{Number of Isolated Items Score IL} & Score capturing the influence of isolated modules on independence level. \\
       
        \midrule
            \multicolumn{2}{l}{\textit{Architecture Anti-pattern Counts}} \\
        \midrule
            \textbf{Clique Count (Total)} & Total instances of strongly connected components. \\
            \textbf{Clique File Count} & Number of files involved in cliques. \\
            \textbf{Unhealthy Inheritance Count (Total)} & Total cases of inheritance-based design smells. \\
            \textbf{Unhealthy Inheritance File Count} & Files affected by inheritance anti-patterns. \\
            \textbf{Package Cycle Count (Total)} & Instances of cyclic dependencies across packages. \\
            \textbf{Package Cycle File Count} & Files implicated in package-level cycles. \\
            \textbf{Total Anti-patterns Count} & Aggregate number of anti-pattern occurrences. \\
            \textbf{Total Anti-pattern Files} & Aggregate file count affected by anti-patterns. \\
        \bottomrule
    \end{tabular}
\end{table}

\textbf{\emph{Class-level and method-level metrics.}} As Java is an object-oriented programming language, we evaluate code quality using the renowned Chidamber and Kemerer (CK) metrics suite. To support this analysis, we employ the CK tool developed by Aniche~\citep{aniche_java_2015}, which computes a comprehensive set of 49 class-level and 30 method-level metrics. Among these are the six canonical CK metrics: Weighted Methods per Class (WMC), Depth of Inheritance Tree (DIT), Number of Children (NOC), Coupling Between Objects (CBO), Response for a Class (RFC), and Lack of Cohesion in Methods (LCOM). A detailed overview of some of the selected class-level metrics is provided in Table~\ref{tab:ck_class_metrics}. Our shared dataset contains the whole list.

\begin{table}
\centering
\caption{Class-Level Metrics and Descriptions.}
\label{tab:ck_class_metrics}
\begin{tabular}{p{3.5cm}p{7.5cm}}
\toprule
\textbf{Metric} & \textbf{Description} \\
\midrule
\textbf{CBO} & Coupling Between Objects: number of classes to which a class is coupled. \\
\textbf{WMC} & Weighted Methods per Class: sum of McCabe complexity of methods. \\
\textbf{DIT} & Depth of Inheritance Tree: depth of class in inheritance hierarchy. \\
\textbf{NOC} & Number of Children: number of immediate subclasses. \\
\textbf{RFC} & Response For a Class: number of unique methods that can be invoked. \\
\textbf{LCOM} & Lack of Cohesion of Methods: dissimilarity of methods in a class. \\
\textbf{TCC} & Tight Class Cohesion: ratio of directly connected method pairs. \\
\textbf{LCC} & Loose Class Cohesion: ratio of direct and indirect method connections. \\
\textbf{Total Methods Quantity} & Total number of methods in a class. \\
\textbf{Static Methods Quantity} & Number of static methods. \\
\textbf{Public Methods Quantity} & Number of public methods. \\
\textbf{Private Methods Quantity} & Number of private methods. \\
\textbf{Protected Methods Quantity} & Number of protected methods. \\
\textbf{Default Methods Quantity} & Number of default (package-private) methods. \\
\textbf{Visible Methods Quantity} & Number of non-private methods. \\
\textbf{Abstract Methods Quantity} & Number of abstract methods. \\
\textbf{Final Methods Quantity} & Number of final methods. \\
\textbf{Total Fields Quantity} & Total number of fields. \\
\textbf{Protected Fields Quantity} & Number of protected fields. \\
\textbf{Default Fields Quantity} & Number of default (package-private) fields. \\
\textbf{Final Fields Quantity} & Number of final fields. \\
\textbf{NOSI} & Number of Static Invocations: calls to static methods. \\
\textbf{LOC} & Lines of Code: non-empty, non-comment lines. \\
\textbf{Return Quantity} & Number of return statements. \\
\textbf{Loop Quantity} & Number of loop statements (for, while, do-while). \\
\textbf{Comparisons Quantity} & Number of comparison operations (==, !=). \\
\textbf{Try-Catch Quantity} & Number of try-catch blocks. \\
\textbf{Parenthesized Expressions Quantity} & Number of expressions in parentheses. \\
\textbf{Assignments Quantity} & Number of assignment operations. \\
\textbf{Math Operations Quantity} & Number of mathematical operations. \\
\textbf{Variables Quantity} & Number of variable declarations. \\
\textbf{Max Nested Blocks Quantity} & Maximum depth of nested blocks. \\
\textbf{Anonymous Classes Quantity} & Number of anonymous inner classes. \\
\textbf{Inner Classes Quantity} & Number of named inner classes. \\
\textbf{Lambdas Quantity} & Number of lambda expressions. \\
\textbf{Unique Words Quantity} & Number of distinct words in source code. \\
\textbf{Modifiers} & Modifiers used in the class and its members. \\
\bottomrule
\end{tabular}
\end{table}

In addition to the 29 method-level metrics provided by the CK tool, we extend the metric set by incorporating an additional metric, Readability, which estimates the cognitive effort required to understand a method based on syntactic and lexical features such as identifier clarity, control structure complexity, and textual coherence. We have used the readability score generation tool proposed by Buse et al.~\citep{buse_learning_2010}. Conversely, we exclude the \textit{hasJavaDoc} metric from our analysis. This metric simply captures whether a method has associated documentation, without offering any insight into the quality or content of that documentation. Also, as we aggregate all method-level features to produce a fixed-length representation for each app—as described later in this section—this binary metric is not suitable. Its boolean nature prevents meaningful aggregation using statistical measures such as minimum, maximum, mean, or percentiles. A sample set of the 29 method-level metrics used in this study is summarized in Table~\ref{tab:ck_method_metrics}.

\begin{table}
\centering
\caption{Method-Level Metrics and Descriptions.}
\label{tab:ck_method_metrics}
\begin{tabular}{p{4.5cm}p{6.5cm}}
\toprule
\textbf{Metric} & \textbf{Description} \\
\midrule
\textbf{Fan-In} & Number of other methods that call this method. \\
\textbf{Fan-Out} & Number of methods this method calls. \\
\textbf{LOC} & Lines of Code: number of lines in the method excluding comments and blanks. \\
\textbf{Return Quantity} & Number of return statements in the method. \\
\textbf{Variables Quantity} & Number of variables declared in the method. \\
\textbf{Parameters Quantity} & Number of parameters the method accepts. \\
\textbf{Methods Invoked Quantity} & Number of distinct methods invoked. \\
\textbf{Methods Invoked Local Quantity} & Number of local methods directly invoked. \\
\textbf{Methods Invoked Indirect Local Quantity} & Number of local methods indirectly invoked. \\
\textbf{Loop Quantity} & Number of loop constructs in the method (for, while, do-while). \\
\textbf{Comparisons Quantity} & Number of comparison operations (==, !=, <, >, etc.). \\
\textbf{Try-Catch Quantity} & Number of try-catch blocks. \\
\textbf{Parenthesized Expressions Quantity} & Number of expressions enclosed in parentheses. \\
\textbf{Assignments Quantity} & Number of assignment statements. \\
\textbf{Math Operations Quantity} & Number of mathematical operations (+, -, *, /, etc.). \\
\textbf{Max Nested Blocks Quantity} & Maximum depth of nested code blocks. \\
\textbf{Lambdas Quantity} & Number of lambda expressions. \\
\textbf{Unique Words Quantity} & Number of unique words used in the method source. \\
\textbf{Modifiers} & Access and non-access modifiers applied to the method. \\
\textbf{Log Statements Quantity} & Number of logging statements. \\
\textbf{Readability} & Readability score indicating ease of understanding the method. \\
\bottomrule
\end{tabular}
\end{table}

While system-level metrics yield a single, fixed set of values per application, class-level and method-level metrics produce variable-sized metric sets depending on the number of classes and methods in each app. However, statistical and machine learning models require a consistent, fixed-length representation across all samples. A common approach to address this involves computing summary statistics such as the minimum, maximum, mean, and median. Yet, these aggregates are often sensitive to outliers, which can distort the underlying distribution.

To capture the structural patterns more robustly and mitigate the influence of extreme values, we compute 11 percentiles for each metric, ranging from the 10$^{th}$ to the 99$^{th}$ percentile (i.e., 10$^{th}$, 20$^{th}$, ..., 90$^{th}$, 95$^{th}$, and 99$^{th}$ percentiles). For class-level metrics, we exclude the features \textit{TCC}, \textit{LCC}, and \textit{LCOM*} due to their high rate of missing values. It is also important to note that the CK tool's class-level metrics are computed over all types of classes, including normal, anonymous, and inner classes.

For normal classes, we compute the minimum, maximum, mean, and 11 percentiles for all class-level features. In addition to these statistics, some metrics carry global significance. We identify seven such important global metrics: \texttt{lambdasQty}, \texttt{totalMethodsQty}, \texttt{innerClassesQty}, \texttt{LOC}, \texttt{anonymousClassesQty}, \texttt{loopQty}, and \texttt{uniqueWordsQty}. For these metrics, we also calculate their total count across the entire codebase for all types of classes to capture their overall impact. To complement the aggregated features, we introduce a new metric, \texttt{TotalNormalClasses}, which captures the number of standard classes in the app. Applications with fewer than five normal classes are excluded (57 applications) from the analysis to ensure statistical reliability.

For method-level metrics, we exclude constructor methods and compute the minimum, maximum, mean, and the same 11 percentiles for each metric. After performing this aggregation process, we obtain a class-level metrics dataset comprising 652 features and a method-level metrics dataset with 406 features per application.

\textbf{\emph{Code Smells.}} Since code smells significantly increase the bug proneness of source code \citep{palomba_diffuseness_2018}, we extracted various types of code smells as features as presented in Table \ref{tab:smells}.

\begin{table}[ht]
\centering
\caption{A Subset of Code Smells and Descriptions.}
\begin{tabular}{p{3cm}p{8cm}}
\hline
\textbf{Metric/Smell} & \textbf{Description} \\
\hline
\textbf{Complex Conditional} & Presence of nested or convoluted conditional logic that reduces code readability and maintainability. \\
\textbf{Complex Method} & A method exhibiting high cyclomatic complexity, making it difficult to test and understand. \\
\textbf{Empty catch clause} & A catch block that does not contain any handling logic, which may hide potential exceptions. \\
\textbf{Long Identifier} & Overly verbose identifier names that can reduce code clarity. \\
\textbf{Long Method} & A method that is excessively long, violating the single-responsibility principle. \\
\textbf{Long Statement} & A single line of code that contains too many operations, reducing readability. \\
\textbf{Magic Number} & Usage of unexplained numeric literals instead of named constants. \\
\textbf{Missing default} & A switch-case structure without a default case, risking unhandled conditions. \\
\textbf{Deficient Encapsulation} & Exposure of internal class details that should be hidden, violating encapsulation principles. \\
\textbf{Insufficient Modularization} & Modules or classes that are too large or do not encapsulate coherent responsibilities. \\
\textbf{Unutilized Abstraction} & Abstract classes or interfaces that are defined but never instantiated or extended. \\
\textbf{Feature Concentration} & Multiple unrelated responsibilities implemented within a single class or module. \\
\textbf{Long Parameter List} & Methods that accept too many parameters, increasing complexity and decreasing reusability. \\
\textbf{Broken Hierarchy} & Inheritance structures that are incorrectly implemented, leading to poor reuse and maintainability. \\
\textbf{Broken Modularization} & Logical modules that are fragmented across packages or files. \\
\textbf{Cyclic-Dependent Modularization} & A cycle of module dependencies, leading to tight coupling and hindered modularity. \\
\textbf{Imperative Abstraction} & Classes that contain detailed implementation rather than abstract behaviors. \\
\textbf{Missing Hierarchy} & Expected inheritance hierarchies are absent, possibly indicating poor reuse. \\
\textbf{Multipath Hierarchy} & A class inherits the same superclass through multiple paths, complicating inheritance resolution. \\
\textbf{Rebellious Hierarchy} & Subclasses that override parent behaviors in ways that break substitutability. \\
\textbf{Unexploited Encapsulation} & Failure to make use of available encapsulation mechanisms, e.g., public fields. \\
\textbf{Unnecessary Abstraction} & Abstractions (interfaces or abstract classes) that do not provide additional value. \\
\textbf{God Component} & A module that centralizes too many responsibilities, similar to a God Class. \\
\hline
\end{tabular}
\label{tab:smells}
\end{table}

There are many tools available for code smell detection, some of which take the APK as input \citep{hecht2015tracking,sonarqube_sonarqube_2023}, while others operate solely on source code \citep{palomba_lightweight_2017,moha_decor_2010,sharma_designite_2016}. For our analysis, we chose the DesigniteJava tool, which is a version of the original Designite tool \citep{sharma_designite_2016} specifically designed for Java. The primary reason for selecting this tool is its comprehensive support for various types of code smells, including architectural, design, implementation, and test smells. For this study, we focused on architectural, design, and implementation smells, which collectively account for a total of 34 code smells.

\textbf{\emph{Other Metadata.}} We identified and collected the following app-specific features: \textit{permissions}, \textit{activity count}, \textit{genre}, and \textit{advertisement support}.

The permissions feature is a list of strings, such as Microphone, Location, Contact, etc. The activity count feature represents the number of Android Activities in an app, and thus, is likely to be correlated with the number of features offered by an app. The genre represents the category of the app, and advertisement support indicates whether the app shows advertisements or not.

The data for permissions, genre, and advertisement support were collected from the Google Play Store. Advertisement support is represented as a binary value (indicating the presence or absence of ads), while genre and permissions are categorical string features. To make these string-based features suitable for machine learning models, we applied one-hot encoding using scikit-learn~\citep{scikit-learn}. This technique transforms each unique category into a separate binary column, where a value of 1 indicates the presence of that category and 0 indicates its absence. This allows models to interpret categorical data without assuming any ordinal relationship between the categories~\citep{zheng2018feature}. The activity count was determined by parsing the \texttt{AndroidManifest.xml} file of each app. 

After processing, we obtained 16 permission features and 40 genre features, resulting in a total of 58 app-specific features for analysis. The features are presented in Table \ref{tab:feature_metadata_full}.

\begin{table}[ht]
\centering
\caption{Metadata Description of Selected App Features.}
\label{tab:feature_metadata_full}
\begin{tabular}{p{2.5cm}p{8.5cm}}
\hline
\textbf{Feature} & \textbf{Description} \\
\hline
\textbf{Genre} & A categorical variable representing the content classification of an application. The possible values include: \textit{ART\_AND\_DESIGN}, \textit{AUTO\_AND\_VEHICLES}, \textit{BOOKS\_AND\_REFERENCE}, \textit{BUSINESS}, \textit{COMICS}, \textit{COMMUNICATION}, \textit{DATING}, \textit{EDUCATION}, \textit{ENTERTAINMENT}, \textit{EVENTS}, \textit{FINANCE}, \textit{FOOD\_AND\_DRINK}, \textit{GAME\_ACTION}, \textit{GAME\_ADVENTURE}, \textit{GAME\_ARCADE}, \textit{GAME\_BOARD}, \textit{GAME\_CARD}, \textit{GAME\_CASINO}, \textit{GAME\_CASUAL}, \textit{GAME\_EDUCATIONAL}, \textit{GAME\_PUZZLE}, \textit{GAME\_RACING}, \textit{GAME\_ROLE\_PLAYING}, \textit{GAME\_SIMULATION}, \textit{GAME\_SPORTS}, \textit{GAME\_STRATEGY}, \textit{GAME\_TRIVIA}, \textit{GAME\_WORD}, \textit{HEALTH\_AND\_FITNESS}, \textit{HOUSE\_AND\_HOME}, \textit{LIBRARIES\_AND\_DEMO}, \textit{LIFESTYLE}, \textit{MAPS\_AND\_NAVIGATION}, \textit{MEDICAL}, \textit{MUSIC\_AND\_AUDIO}, \textit{NEWS\_AND\_MAGAZINES}, \textit{PARENTING}, \textit{PERSONALIZATION}, \textit{PHOTOGRAPHY}, \textit{PRODUCTIVITY}, \textit{SHOPPING}, \textit{SOCIAL}, \textit{SPORTS}, \textit{TOOLS}, \textit{TRAVEL\_AND\_LOCAL}, \textit{VIDEO\_PLAYERS}, and \textit{WEATHER}. \\
\hline
\textbf{Uses Permission} & A multi-label categorical variable indicating system-level permissions requested by the application. The enumerated values are: \textit{Location}, \textit{Phone}, \textit{Photos/Media/Files}, \textit{Storage}, \textit{Wi-Fi connection information}, \textit{Device ID \& call information}, \textit{Other}, \textit{Uncategorized}, \textit{Camera}, \textit{Microphone}, \textit{Identity}, \textit{Calendar}, \textit{Contacts}, \textit{Device \& app history}, \textit{SMS}, and \textit{Wearable sensors/Activity data}. These permissions are indicative of the app's access scope and potential privacy implications. \\
\hline
\textbf{Advertisement Support} & A binary feature indicating whether the application contains embedded advertising services. \\
\hline
\textbf{Activity Count} & A numerical feature capturing the total number of declared Android activity components in the app’s manifest file.\\
\hline
\end{tabular}
\end{table}

\subsection{Feature Selection}
\label{sec:feature_selection}
After excluding 57 applications from the initial pool of 503 during class-level code metrics extraction, we obtained a final dataset of 446 Android applications, each characterized by 1,167 features. While this comprehensive feature set was designed to capture a wide range of structural and semantic properties of software systems, the inclusion of high-dimensional feature spaces can adversely affect the performance of machine learning models—a phenomenon commonly attributed to the curse of dimensionality. To mitigate this issue and enhance model generalizability, we applied feature selection techniques to reduce the dimensionality of the dataset while preserving relevant information. The specific subset of features employed in each experimental setting, along with their respective contributions to model performance, is discussed in detail in Section~\ref{sec:results}. 

\textbf{\emph{Feature Set 1: Size-only.}}
Although numerous software metrics have been proposed and empirically evaluated as predictors of maintainability and other quality attributes, the reliability of these metrics, beyond basic size, remains a topic of ongoing debate~\citep{chowdhury_revisiting_2022,gil_correlation_2017,shepperd_critique_1988,gil_when_2016}. Several studies argue that the size of code is the most robust and consistent predictor of maintenance effort~\citep{el_emam_confounding_2001,gil_correlation_2017}, while many other code metrics exhibit strong collinearity with size~\citep{herraiz_towards_2007}. Based on this evidence, we constructed a minimal baseline model using only a size-related feature to assess its standalone predictive power. To measure size, we used the \textit{CLOC}~\citep{adanial_cloc} tool, which analyzes a project directory and reports the number of blank lines, comment lines, and lines of code. For each app, we computed the lines of code as size, excluding blank lines and comments.

\textbf{\emph{Feature Set 2: Handpicked.}}
To construct a compact yet representative subset of features, we manually selected 28 features based on domain knowledge and prior empirical studies. The first selected feature is \textit{size}, given its well-documented predictive value~\citep{el_emam_confounding_2001,gil_correlation_2017,herraiz_towards_2007}. Additionally, we included two system-level features: \textit{decouplingLevel}, which quantifies the extent to which a software system adheres to modular design principles~\citep{mo_decoupling_2016}, and \textit{total\_InstanceCount}, representing the cumulative count of detected anti-patterns across the codebase. Together, these features offer a coarse-grained overview of architectural quality.

At the class level, we selected three canonical metrics from the CK suite: Coupling Between Objects (CBO), Weighted Methods per Class (WMC), and Response for a Class (RFC). These metrics have demonstrated consistent relevance in empirical studies evaluating code quality and defect proneness~\citep{singh2010empirical,shatnawi2008effectiveness,olague2007empirical,radjenovic_software_2013}. As adding all the 11 percentiles for these three metrics would increase the feature set size significantly, we computed only the 10th, 50th, and 90th percentiles, resulting in a total of nine class-level features.

From the method-level metrics, we selected four: \textit{WMC}, \textit{fanin}, \textit{fanout}, and \textit{readability}. Percentiles (10th, 50th, and 90th) were again computed for each, producing 12 method-level features in total. The \textit{readability} metric is of particular interest as it estimates the cognitive load required to comprehend a method, based on syntactic and lexical attributes.

In addition to these metrics, we incorporated three widely studied code smells: \textit{God Components}, \textit{Long Statement}, and \textit{Long Method}, which have been shown to correlate with software defects and maintainability issues~\citep{palomba_smells_2016,catolino_improving_2020,palomba_toward_2019}. Finally, we included the \textit{containsAds} feature, a binary indicator of whether the application integrates advertising-related code.

This manually curated feature set balances interpretability and predictive relevance, leveraging insights from prior literature and expert judgment to represent core aspects of software complexity, modularity, and maintainability.

\textbf{\emph{Feature Set 3: Voting.}}
In contrast to the previous two feature sets, we also add another feature set with a more systematic approach by applying feature selection algorithms and voting. 
Feature selection algorithms can be broadly categorized into filter, wrapper, embedded, hybrid, and ensemble model-based methods \citep{barbieri_analysis_2024}. Each of these techniques has its strengths and weaknesses. Li et al. \citep{li_recent_2017} described various scenarios and the appropriate feature selection techniques for each. They recommended an ensemble approach combining multiple types of feature selection techniques for datasets with high dimensionality and small sample sizes. Following their guidance, we employed a combination of feature selection methods, including two filter-type algorithms, one wrapper algorithm, and three embedded-type algorithms.

For classification and regression tasks, the choice of feature selection algorithms was tailored to address the unique characteristics of each problem. For classification, the algorithms were categorized into three types: filter methods, which included Pearson Correlation and Chi2 Correlation; wrapper methods, represented by the Support Vector Classifier; and embedded methods, consisting of Logistic Regression, Random Forest Classifier, and Light Gradient Boosting Classifier. For regression tasks, a similar categorization was followed. The filter methods included Pearson Correlation and Analysis of Variance, while the wrapper approach utilized the Support Vector Regressor. The embedded methods comprised Lasso Regressor, Ridge Regressor, and Random Forest Regressor.

Each of the feature selection algorithms provides a ranked list of $n$ features, where $n$ represents the number of features selected by the algorithm. For our experiments, we set $n$ to 25, allowing each algorithm to suggest its top 25 features. To aggregate the features suggested by different algorithms, we apply a voting mechanism. Specifically, we select only those features that are recommended by at least 50\% of the algorithms, which in our case corresponds to three feature selection algorithms out of the six used. This aggregation ensures that the final selected features are robust and agreed upon by multiple algorithms, reducing the likelihood of including irrelevant or redundant features.

\subsection{Learning Algorithms and Evaluation Metrics}
\label{sec:methodology_model}
We employed a set of popular machine-learning algorithms for both classification and regression tasks. For the classification task, we employed a diverse set of algorithms encompassing both simple and complex models. Specifically, we utilized Logistic Regression (LR), Decision Tree Classifier (DT), Random Forest Classifier (RF), Gradient Boosting Classifier (GB), and Multi-Layer Perceptron (MLP). This selection spans a spectrum of model complexities, ranging from simple and linear approaches such as LR to highly non-linear and complex models like MLP.  To evaluate the performance of these models, we calculated class-wise metrics, including precision, recall, F1 score, area under the curve (AUC), and Matthew's Correlation Coefficient (MCC).

For the regression tasks, we employed a diverse suite of models, including Lasso, Ridge, Decision Tree Regressor (DT), Random Forest Regressor (RF), Gradient Boosting Regressor (GB), and Multi-Layer Perceptron Regressor (MLP). The selection strategy mirrors that of the classification setting, encompassing both linear models (e.g., Lasso, Ridge) and more complex non-linear estimators (e.g., MLP). The regression models were evaluated using metrics such as root mean squared error (RMSE), mean absolute error (MAE), and the coefficient of determination ($R^2$ score).

To ensure robust and unbiased model evaluation, we employed the Leave-One-Out Cross-Validation (LOOCV) technique. LOOCV is a reliable method for small datasets, where each instance serves as a test set exactly once while the remaining instances are used for training. By aggregating the results across all iterations, we obtained a comprehensive understanding of the models' generalization performance.

\section{APPROACH, ANALYSIS, AND RESULTS}
\label{sec:results}
In this section, we present the approach used to address each of the three research questions, along with the corresponding findings.

\subsection{\textbf{RQ1:} To what extent can internal software metrics predict an app’s user rating?}
To address this research question, we first examined the feasibility of predicting actual app ratings. Next, we categorized the apps into two groups, \emph{Popular} and \emph{Unpopular}, based on their ratings, and investigated the feasibility of predicting an app's popularity. To achieve this, we conducted a series of experiments using regression and classification models.
\subsubsection{\textbf{Regression: Predicting the Actual Ratings}}
User ratings for mobile applications are generally provided on a discrete scale ranging from 1 to 5. To enable a continuous prediction task, we computed the mean rating for each application based on all associated user reviews. Applications with no user reviews are excluded from the analysis, as they cannot have a valid rating assigned (ratings cannot be less than 1). After removing these 39 applications, our regression analysis is conducted using the remaining 407 applications. 

As outlined in Section~\ref{sec:feature_selection}, we considered three distinct feature sets for our analysis: \texttt{Size-only}, \texttt{Handpicked}, and \texttt{Voting}. Among these, the \texttt{Voting} approach to feature selection identified three features as most predictive of user ratings: \texttt{genreId\_GAME\_BOARD}, \texttt{fanin\_30\_class}, \texttt{logStatementsQty\_max\_class}. For example, \texttt{fanin\_30\_class} is the 30$^{th}$ percentile of the metric \texttt{fanin} at the class level granularity.

Building upon these three feature sets, we conducted three groups of experiments in which six regression models were applied to predict the continuous app ratings: \texttt{MLPRegressor} (MLP), \texttt{Lasso}, \texttt{Ridge}, \texttt{DecisionTreeRegressor} (DT), \texttt{RandomForestRegressor} (RF), and \texttt{GradientBoostingRegressor} (GB). To ensure that the predicted ratings remained within the valid interval [1,5], we applied a post-processing step wherein predictions falling below 1 were clamped to 1, and those exceeding 5 were truncated to 5. This normalization was necessary due to the unbounded nature of regression outputs, which can produce values outside the expected domain. The comparative results of the regression models are summarized in Table~\ref{tab:rating_regression}.

Overall, the results indicate a clear trend: models trained on the \texttt{Voting} feature set consistently outperform those trained on \texttt{Size-only} and \texttt{Handpicked} features. Models trained using the \texttt{Size-only} feature set generally performed worse, with higher RMSE and MAE values and $R^2$ values that are below zero. This underscores the limited predictive power of size-related metrics alone and highlights the added value of incorporating domain-informed or statistically selected features. Among the evaluated models, Ridge regression achieved the best overall performance using the \texttt{Voting} features, with an RMSE of 0.69, MAE of 0.51, and the highest $R^2$ value of 0.05. In contrast, the MLPRegressor demonstrated the weakest performance.

\begin{table*}[ht]
\centering
\caption{Regression metrics (RMSE, MAE, $R^2$) of rating regressors evaluated using three feature selection strategies: Size-only, Handpicked, and Voting. Bold values indicate the best metric scores within each feature set.}
\label{tab:rating_regression}
\resizebox{\textwidth}{!}{
\begin{tabular}{lccc ccc ccc}
\toprule
\multirow{2}{*}{\textbf{Model}} & 
\multicolumn{3}{c}{\textbf{Size-only}} & 
\multicolumn{3}{c}{\textbf{Handpicked}} & 
\multicolumn{3}{c}{\textbf{Voting}} \\
\cmidrule(lr){2-4} \cmidrule(lr){5-7} \cmidrule(lr){8-10}
& \textbf{RMSE} & \textbf{MAE} & \textbf{$R^2$}
& \textbf{RMSE} & \textbf{MAE} & \textbf{$R^2$}
& \textbf{RMSE} & \textbf{MAE} & \textbf{$R^2$} \\
\midrule
MLP   & 3.00 & 2.90 & -16.92 & 2.10 & 1.76 & -7.82 & 0.73 & 0.56 & -0.05 \\
Lasso & \textbf{0.71} & \textbf{0.53} & \textbf{-0.01} & \textbf{0.71} & \textbf{0.53} & \textbf{-0.01} & 0.71 & 0.53 & 0.00 \\
Ridge & \textbf{0.71} & \textbf{0.53} & \textbf{-0.01} & 0.72 & 0.54 & -0.03 & \textbf{0.69} & \textbf{0.51} & \textbf{0.05} \\
DT    & 1.05 & 0.78 & -1.20 & 1.01 & 0.76 & -1.04 & 0.73 & 0.55 & -0.07 \\
RF    & 0.89 & 0.67 & -0.59 & 0.72 & 0.55 & -0.03 & 0.72 & 0.54 & -0.03 \\
GB    & 0.80 & 0.58 & -0.26 & 0.75 & 0.57 & -0.12 & 0.71 & 0.53 & 0.00 \\
\bottomrule
\end{tabular}
}
\end{table*}

A closer examination of the regression outcomes reveals that, for the majority of cases, the RMSE remains below 1 and the MAE is under 0.6.  These values may suggest that the internal code metrics possess reasonable predictive capability: for instance, the Ridge regression model achieves an MAE of 0.51, indicating that it can predict application ratings with around half-point deviation on average. Such results imply that, in absolute terms, the models are able to approximate user ratings with relatively low error. 

However, this positive impression is tempered by a contrasting trend in the $R^2$ scores. While RMSE and MAE reflect the magnitude of prediction errors, the $R^2$ metric evaluates the proportion of variance in the target variable that is explained by the model. While $R^2$ values close to 1 indicate strong explanatory power, the metric can range from $-\infty$ to 1. A value of 0 means the model performs no better than predicting the mean of the target variable, while negative values indicate that the model performs worse than this baseline. In our experiments, most models yield $R^2$ scores that are near zero or even negative, suggesting limited ability to capture the variance in user ratings. Notably, the Ridge model—despite its low error metrics—achieves only a modest $R^2$ of 0.05, and several models (e.g., MLP) produce significantly negative values.

\begin{figure}[ht]
  \centering
  \subfloat[CDF of actual ratings.\label{fig:cdf_rating}]{
    \includegraphics[width=0.7\textwidth]{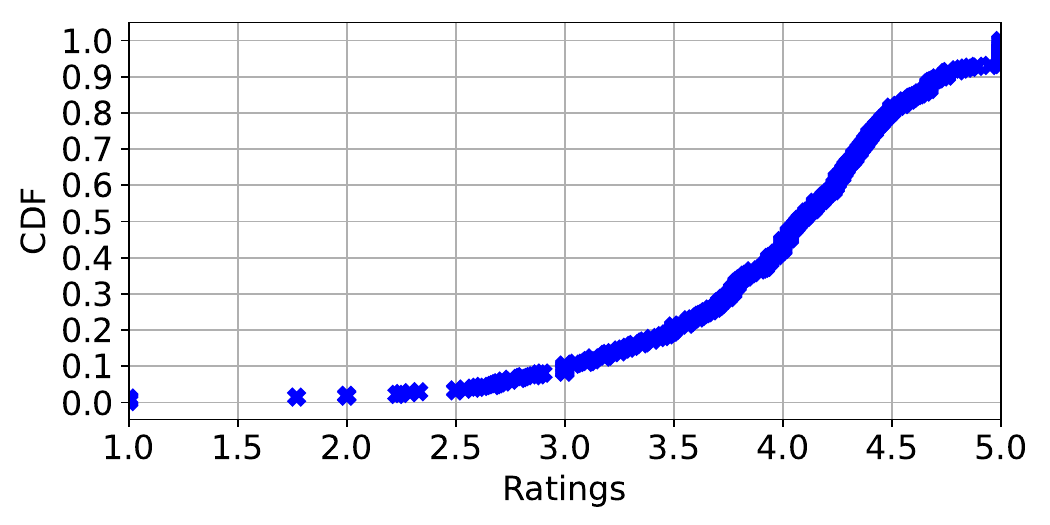}
  }
  \hfill
  \subfloat[CDF of predicted ratings of Ridge regression model.\label{fig:cdf_rating_ridge}]{
    \includegraphics[width=0.7\textwidth]{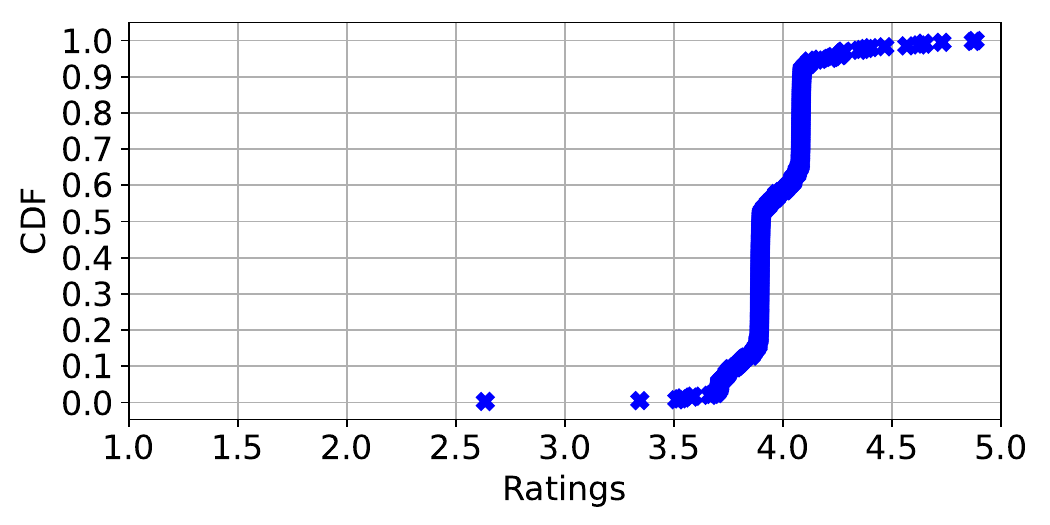}
  }
  \caption{Cumulative Distribution Functions (CDFs) of actual and predicted ratings. Approximately 70\% of the actual ratings fall between 3.0 and 4.5, while over 95\% of the predicted ratings lie within this range, indicating the model's bias towards predicting around the mean value.}
  \label{fig:true_and_predict_cdf}
\end{figure}

To better understand the limited performance, we analyzed the distribution of both actual and predicted ratings using the Cumulative Distribution Function (CDF), as shown in Figure~\ref{fig:true_and_predict_cdf}. The CDF of the true application ratings (Figure~\ref{fig:cdf_rating}) reveals a significant skew in the dataset: approximately 10\% of applications received rating 3, while over 5\% achieved the maximum rating (5). Most notably, nearly 70\% of the applications fall within the 3.0 to 4.5 range, indicating a concentrated distribution around moderately high ratings. This skew is reflected in the predicted values generated by the Ridge regression model (Figure~\ref{fig:cdf_rating_ridge}), where more than 95\% of predictions also lie within the 3.0 to 4.5 interval.

\begin{figure}[ht]
    \centering
    \subfloat[CDF of absolute errors.\label{fig:ridge_rating_cdf_absolute_error}]{
        \includegraphics[width=0.7\textwidth]{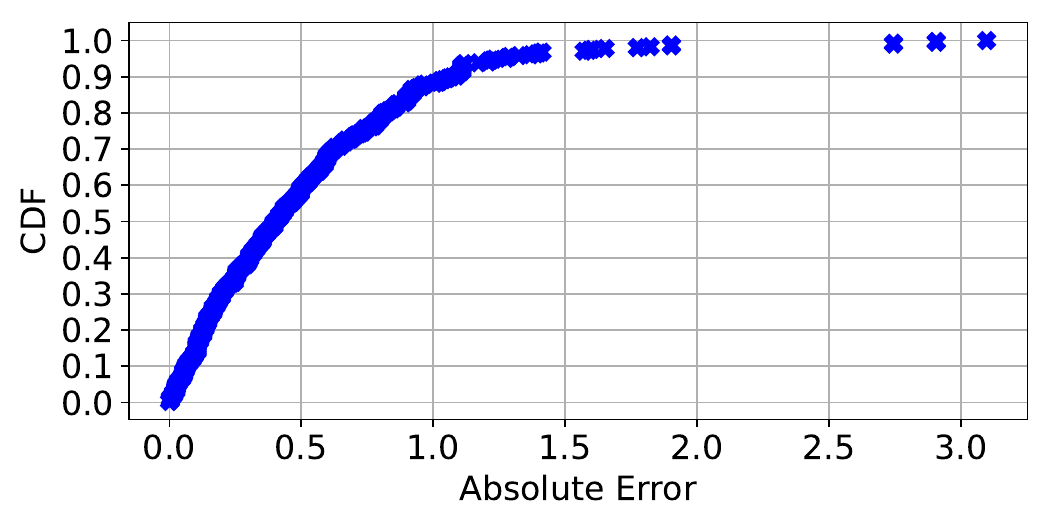}
    }
    \hfill
    \subfloat[CDF of squared errors.\label{fig:ridge_rating_cdf_squared_error}]{
        \includegraphics[width=0.7\textwidth]{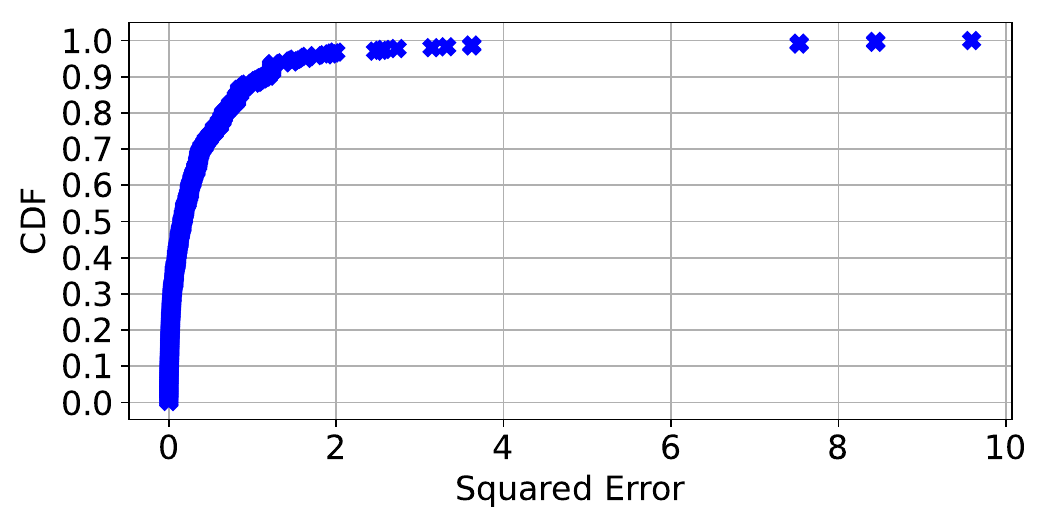}
    }
  \caption{Cumulative Distribution Functions (CDFs) of absolute error and squared error for the Ridge regression model in predicting actual ratings. The plots indicate that a small number of data points account for the majority of the large errors.}
  \label{fig:rating_ridge}
\end{figure}

The model exhibits a clear bias toward the dominant region of the target distribution, which contributes to the relatively low RMSE and MAE values, as illustrated in Figure~\ref{fig:rating_ridge}. Approximately 90\% of the absolute prediction errors fall below 1.0, aligning with the observed Mean Absolute Error (MAE) of 0.51. However, a small subset of instances incurs substantially higher errors, likely corresponding to outliers whose true ratings deviate significantly from the distribution's central mass. Notably, while 60\% of the original ratings fell between 3.5 and 4.5, the model predicted approximately 95\% of applications within this narrow range, indicating a strong centralizing tendency. This behavior reflects the model’s bias towards the most frequent rating region in the training data, causing predictions to cluster around the global mean rather than capturing the full range of rating variation.


\begin{figure}[ht]
    \centering
    \includegraphics[width=1\linewidth]{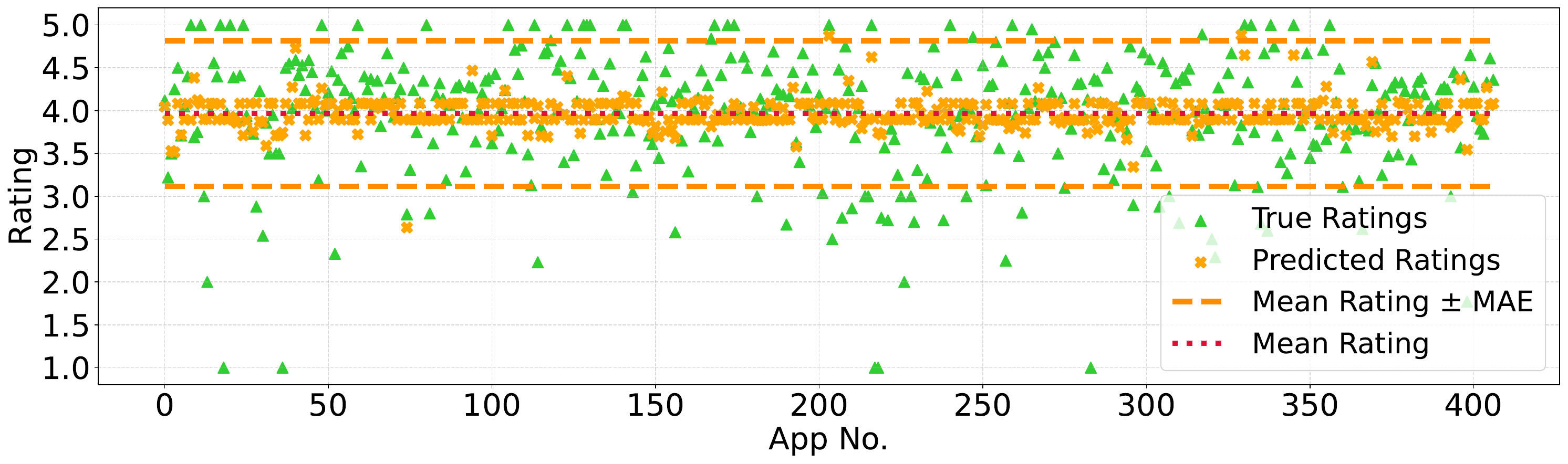}
    \caption{Scatter plot of true ratings and predicted ratings for each app.}
    \label{fig:rating_regression_scatter_line_plot}
\end{figure}

To further verify this, the predicted ratings were plotted against the true ratings in a scatter plot, as shown in Figure~\ref{fig:rating_regression_scatter_line_plot}. Additionally, we included reference lines corresponding to the mean true rating and the bounds defined by $\pm$ the mean absolute error. Most predictions clustered closely around the mean rating line, and nearly all predicted values fell within the $\pm$ MAE bounds, reinforcing the model's tendency to predict values near the global average.

\begin{figure}[ht]
    \centering
    \includegraphics[width=0.55\linewidth]{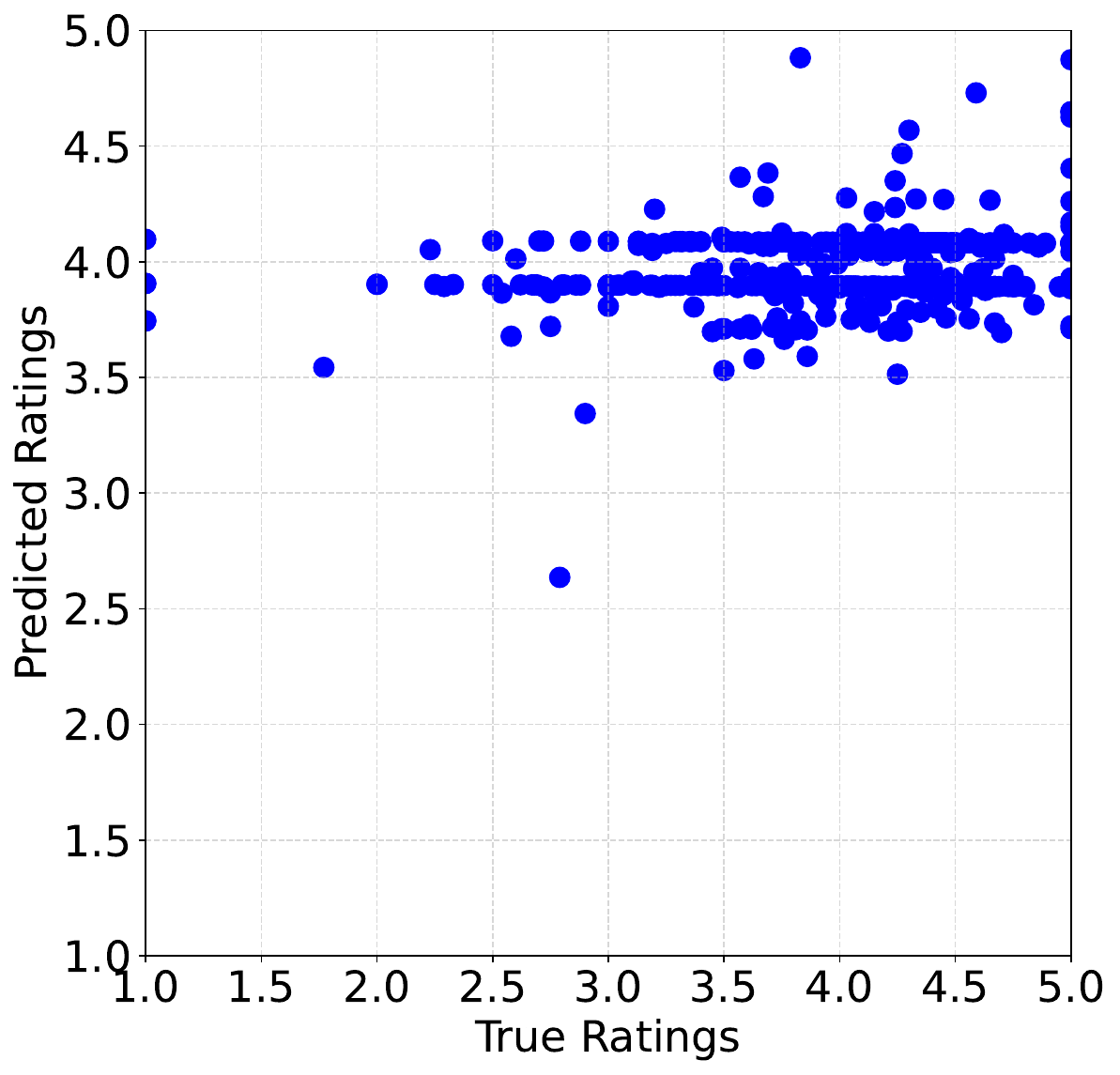}
    \caption{Scatter Plot of True Ratings vs Predicted Ratings.}
    \label{fig:rating_regression_scatter_vs_true_plot}
\end{figure}

Finally, we assessed the alignment between true and predicted ratings using a standard scatter plot, with the actual ratings on the x-axis and the predicted ratings on the y-axis (Figure~\ref{fig:rating_regression_scatter_vs_true_plot}). Ideally, points should concentrate along the diagonal line for an accurate predictive model. However, our results show that predicted ratings were predominantly concentrated between 3.5 and 4.5.

In summary, the strong skew in the dataset leads the regression models to consistently predict values near the mean, resulting in low RMSE and MAE but offering limited insight into the underlying patterns of the data. This behavior is reflected in the uniformly low—often negative—$R^2$ values, indicating that the models fail to meaningfully capture the variance in user ratings. Consequently, the regression results do not offer conclusive evidence regarding the predictive utility of internal code metrics; they neither confirm their effectiveness nor definitively refute their relevance.

To address the inherent bias introduced by the skewed distribution, we propose an alternative approach: reframing the problem as a classification task by discretizing the continuous rating scores into categorical classes. This transformation may help mitigate the centralizing effect observed in regression and enable the models to better distinguish between applications of varying quality. In the following set of experiments, we explore this classification-based formulation to evaluate its potential in uncovering relationships between internal metrics and ratings.
\subsubsection{\textbf{Classification: Predicting Popularity Based on Ratings.}}
We began by categorizing apps into \emph{Popular} and \emph{Unpopular} based on user ratings. Apps with ratings $\geq 3.5$ were classified as \emph{Popular}, while those with ratings below 3.5 were labeled as \emph{Unpopular}, following thresholds established in prior studies \citep{khalid_what_2015,catolino_does_2018}. In this experiment, we retained applications with no user reviews, as the absence of reviews strongly indicates that these applications are indeed \emph{Unpopular}.  This classification resulted in a dataset of 327 \emph{Popular} apps and 119 \emph{Unpopular} apps, presenting a noticeable class imbalance.

Similar to the previous experiments, we also have three distinct feature sets for the classification task: \texttt{Size-only}, \texttt{Handpicked}, and \texttt{Voting}. The \texttt{Voting} method highlighted five key features: \texttt{containsAds}, \texttt{innerClassesQty\_50\_class}, \texttt{genreId\_TRAVEL\_AND\_LOCAL}, \texttt{parametersQty\_20}, and \texttt{parametersQty\_99}. These feature sets were used to train and evaluate various classification models using Leave-One-Out Cross-Validation (LOOCV), as outlined in Section \ref{sec:methodology_model}. 

To address the inherent class imbalance in our dataset, we implemented three widely used resampling techniques: Random Over Sampling (ROS), Random Under Sampling (RUS), and Synthetic Minority Oversampling Technique (SMOTE) \citep{chawla_smote_2002}. These techniques were applied to ensure that the classifier had a more balanced representation of different classes, ultimately improving model generalization. Given that we employed LOOCV as our validation strategy, resampling was performed within each fold to prevent data leakage and maintain the integrity of the training process. This ensured that the model was evaluated on unseen data without exposure to artificially generated or resampled instances from the test fold. Additionally, the numerical features were normalized by applying standard scaling using scikit-learn \citep{scikit-learn}, which standardizes each feature by removing its mean and scaling it to unit variance. This normalization is particularly beneficial for models that rely on gradient-based optimization, such as neural networks.

The results of our classification experiments, presented in Table~\ref{tab:bin_rating_imb_model_performance}, mirror trends observed in the regression setting: models trained on the \texttt{Voting} feature set consistently outperform those utilizing either the \texttt{Size-only} or \texttt{Handpicked} feature sets. In particular, models trained solely on the \texttt{Size-only} features exhibit the weakest performance, with MCC values that are close to or below zero. This indicates poor discriminative ability, often comparable to or worse than random guessing. 

\begin{table}[ht]
\centering
\caption{Performance metrics (Macro F1, MCC, AUC) of rating-based popularity classifiers evaluated using three feature selection strategies: Size-only, Handpicked, and Voting. Bold values indicate the best metric scores within each feature set.}
\label{tab:bin_rating_imb_model_performance}
\resizebox{\textwidth}{!}{
\begin{tabular}{lccccccccc}
\toprule
\multirow{2}{*}{\textbf{Model}} & 
\multicolumn{3}{c}{\textbf{Size-only}} & 
\multicolumn{3}{c}{\textbf{Handpicked}} & 
\multicolumn{3}{c}{\textbf{Voting}} \\
\cmidrule(lr){2-4} \cmidrule(lr){5-7} \cmidrule(lr){8-10}
& \textbf{Macro F1} & \textbf{MCC} & \textbf{AUC}
& \textbf{Macro F1} & \textbf{MCC} & \textbf{AUC}
& \textbf{Macro F1} & \textbf{MCC} & \textbf{AUC} \\
\midrule
MLP & 0.42 & 0.00 & 0.49 & 0.56 & 0.12 & 0.60 & 0.58 & 0.27 & 0.63 \\
LR & 0.42 & 0.00 & 0.37 & 0.58 & 0.18 & 0.63 & 0.56 & 0.23 & 0.62 \\
DT & 0.51 & 0.02 & 0.51 & 0.53 & 0.07 & 0.54 & 0.55 & 0.15 & 0.56 \\
RF & 0.52 & 0.04 & 0.52 & 0.53 & 0.17 & 0.57 & 0.55 & 0.16 & 0.57 \\
GB & 0.52 & 0.09 & 0.51 & 0.56 & 0.19 & 0.55 & 0.56 & 0.21 & 0.56 \\
\midrule
MLP+SMOTE & 0.50 & 0.16 & \textbf{0.63} & 0.59 & 0.18 & 0.62 & \textbf{0.72} & \textbf{0.45} & \textbf{0.72} \\
LR+SMOTE & 0.41 & 0.10 & 0.60 & 0.60 & 0.21 & \textbf{0.64} & 0.55 & 0.20 & 0.73 \\
DT+SMOTE & 0.46 & -0.04 & 0.48 & 0.51 & 0.03 & 0.52 & 0.50 & 0.07 & 0.63 \\
RF+SMOTE & 0.46 & -0.04 & 0.50 & 0.59 & 0.18 & 0.59 & 0.52 & 0.11 & 0.64 \\
GB+SMOTE & 0.54 & 0.09 & 0.52 & 0.58 & 0.16 & 0.57 & 0.51 & 0.13 & 0.66 \\
\midrule
MLP+RUS & 0.50 & 0.14 & \textbf{0.63} & 0.56 & 0.18 & 0.62 & 0.56 & 0.13 & 0.61 \\
LR+RUS & 0.40 & 0.09 & 0.65 & 0.55 & 0.15 & 0.59 & 0.55 & 0.21 & 0.66 \\
DT+RUS & 0.50 & 0.07 & 0.54 & 0.46 & -0.02 & 0.49 & 0.56 & 0.13 & 0.61 \\
RF+RUS & 0.50 & 0.06 & 0.54 & 0.52 & 0.10 & 0.58 & 0.68 & 0.39 & 0.65 \\
GB+RUS & 0.51 & 0.05 & 0.54 & 0.51 & 0.06 & 0.55 & 0.66 & 0.37 & 0.66 \\
\midrule
MLP+ROS & 0.51 & \textbf{0.16} & 0.57 & 0.58 & 0.16 & 0.61 & 0.50 & 0.04 & 0.65 \\
LR+ROS & 0.40 & 0.08 & 0.51 & \textbf{0.61} & \textbf{0.24} & \textbf{0.64} & 0.55 & 0.20 & 0.71 \\
DT+ROS & 0.51 & 0.02 & 0.51 & 0.56 & 0.11 & 0.56 & 0.47 & -0.00 & 0.57 \\
RF+ROS & 0.51 & 0.02 & 0.51 & 0.55 & 0.14 & 0.58 & 0.47 & -0.02 & 0.59 \\
GB+ROS & \textbf{0.55} & 0.10 & 0.53 & 0.56 & 0.12 & 0.55 & 0.48 & 0.01 & 0.58 \\
\bottomrule
\end{tabular}
}
\end{table}

Among the resampling strategies evaluated, the combination of \texttt{MLPClassifier} with Synthetic Minority Oversampling Technique (SMOTE) yielded the most substantial performance gains. Compared to other approaches, such as Random Oversampling (ROS) and Random Undersampling (RUS), SMOTE proved more effective by generating synthetic instances of the minority class rather than merely duplicating existing samples or removing majority class instances. This synthetic data augmentation enabled the classifier to better approximate the decision boundary between classes, thereby enhancing the model’s generalization and robustness.

Overall, the \texttt{MLPClassifier} trained with SMOTE on the \texttt{Voting} feature set achieved the strongest and most balanced performance across all evaluated metrics. The model attained a Macro F1 score of 0.72, indicating a balanced classification performance across both classes. Furthermore, the MCC increased to 0.45, indicating a meaningful improvement in overall predictive reliability and suggesting that the comparatively high F1 score was not due to chance. The Area Under the Receiver Operating Characteristic Curve (AUC-ROC) also reached 0.72, suggesting that the model was more capable of distinguishing between the positive and negative classes.

\begin{figure}[ht]
    \centering
    \includegraphics[width=\textwidth]{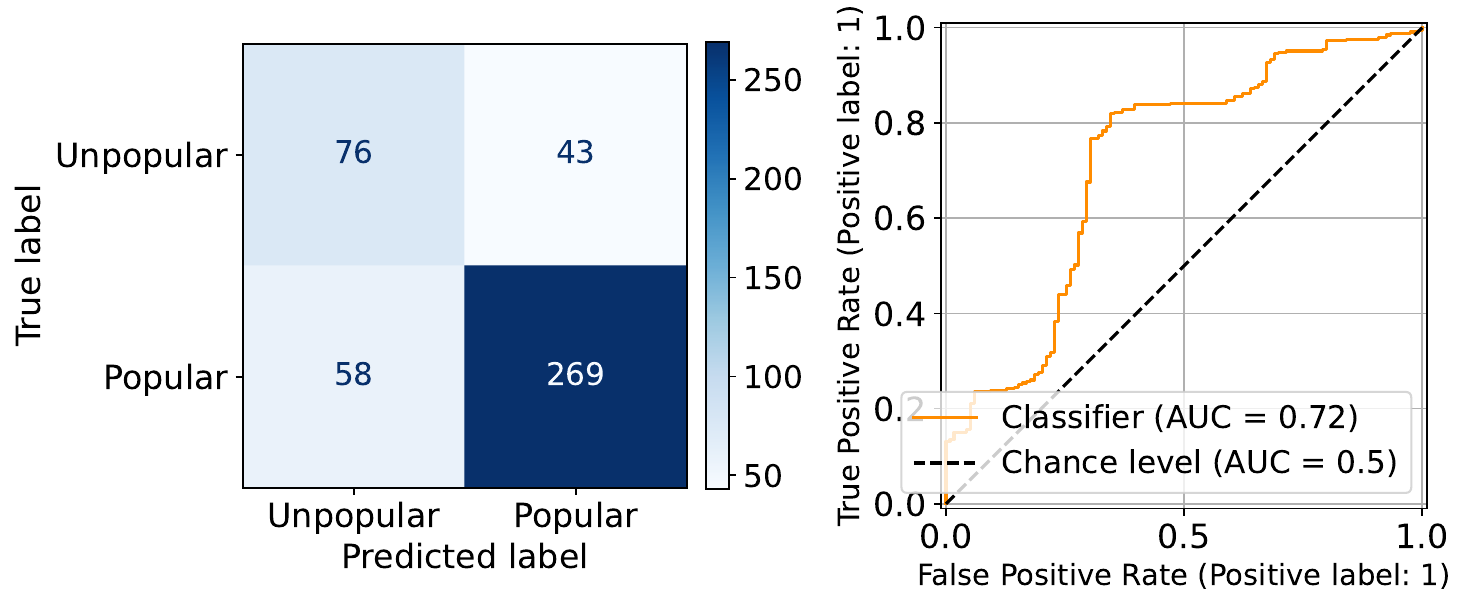}
    \caption{Confusion matrix and AUC-ROC curve for rating-based popularity prediction using MLP classifier with SMOTE.}
    \label{fig:bin_rating_mlp_smote}
\end{figure}

The class-wise performance of the SMOTE-enhanced MLPClassifier, shown in Table \ref{tab:classification_report_bin_rating_mlp_smote}, highlights its effectiveness in mitigating class imbalance. The model achieved an F1 score of 0.601 for the \emph{Unpopular} class, indicating improved minority class recognition while still facing some misclassification challenges. Meanwhile, the \emph{Popular} class demonstrated stronger performance with a precision of 0.862, a recall of 0.823, and an F1 score of 0.842, ensuring high reliability in identifying high-rated apps. These results confirm that SMOTE effectively improved recall for the minority class while maintaining high precision for the majority class, leading to a Macro F1 score of 0.72 and MCC of 0.45. By generating synthetic examples, SMOTE allowed the model to learn better decision boundaries, enhancing overall classification performance. The confusion matrix and AUC-ROC curve in Figure \ref{fig:bin_rating_mlp_smote} further illustrate the classifier’s improved ability to separate \emph{Popular} from \emph{Unpopular} apps, as reflected in a higher true positive rate for the minority class and an AUC score of 0.72.

\begin{table}[ht]
\centering
\caption{Classification performance for rating-based popularity prediction using MLP classifier with SMOTE.}
\label{tab:classification_report_bin_rating_mlp_smote}
\begin{tabular}{lccc}
\hline
\textbf{Class}      & \textbf{Precision} & \textbf{Recall} & \textbf{F1-Score} \\
\hline
Unpopular       & 0.567              & 0.639           & 0.601             \\
Popular         & 0.862              & 0.823           & 0.842             \\
\hline
\end{tabular}
\end{table}

While the classification results are encouraging, they are not perfect. The findings suggest that code metrics contribute meaningfully to predicting application popularity, but they are not sufficient in isolation to achieve optimal performance. The moderate classification scores indicate that, although useful, these features capture only a portion of the factors influencing user perception and engagement.

\begin{summarybox}{Summary of RQ1}
   Due to the skewed data, the regression models did not learn much and made predictions around the mean only. As a result, we were unable to draw meaningful conclusions about the usefulness of internal metrics from the regression results alone. In contrast, classification models---especially \texttt{MLPClassifier} with SMOTE---demonstrated significantly better performance in distinguishing between \emph{Popular} and \emph{Unpopular} apps, highlighting the predictive value of internal software metrics when the problem is appropriately framed.
\end{summarybox}

\subsection{\textbf{RQ2:} To what extent can internal software metrics predict an app’s number of downloads?}
\subsubsection{\textbf{Regression: Predicting the Actual \texttt{DownloadsPerYear}.}}
\begin{figure}[ht]
    \centering
    \includegraphics[width=0.75\linewidth]{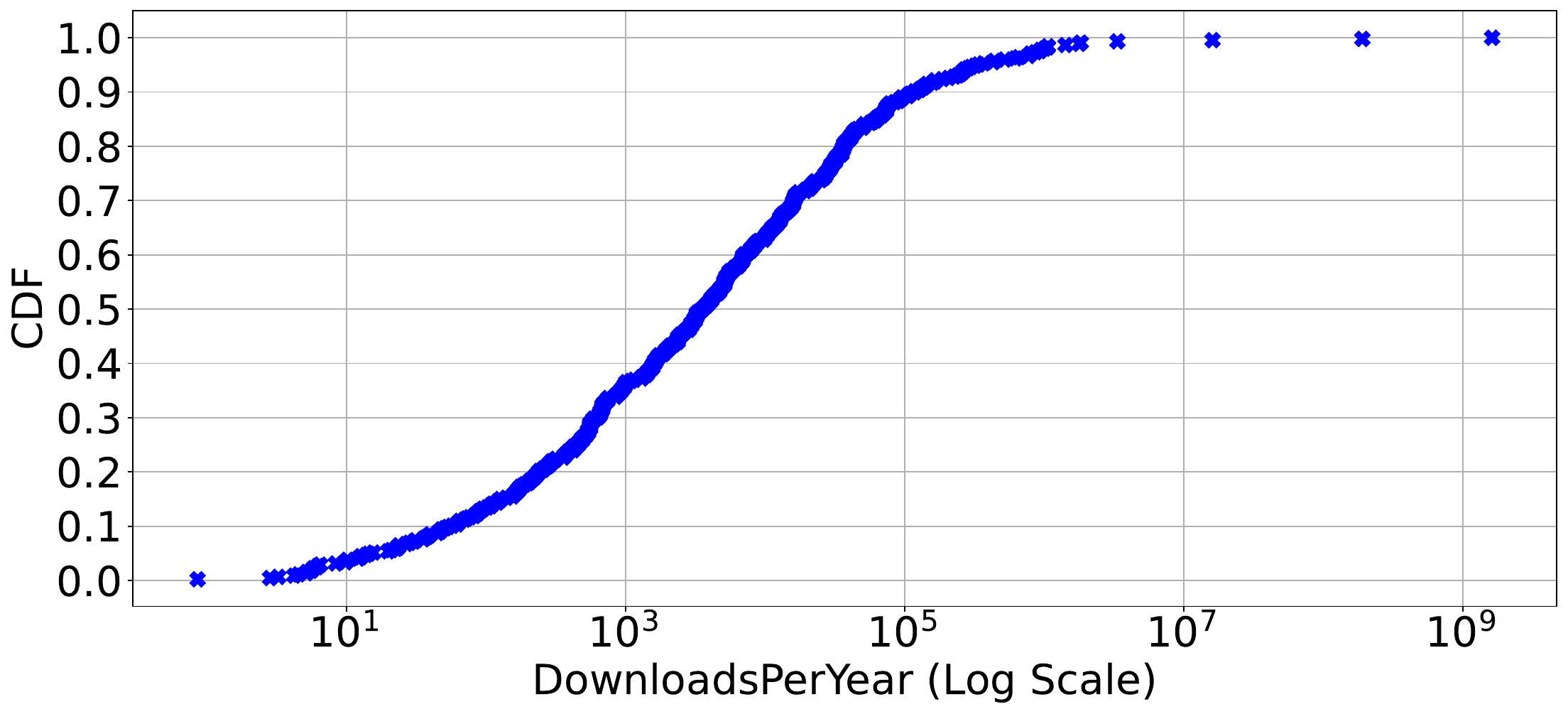}
    \caption{CDF of \texttt{DownloadsPerYear}.}
    \label{fig:cdf_dpy}
\end{figure}

As we explained in Section \ref{sec:popularity_indicators}, the \texttt{DownloadsPerYear} measure was created by normalizing the age impact on total download count. In Figure \ref{fig:cdf_dpy}, we see the CDF of \texttt{DownloadsPerYear} in the log scale, where some values are too high and can be treated as outliers. Removing these outliers will help us to understand the distribution better. 

\begin{figure}
    \centering
    \includegraphics[width=0.75\linewidth]{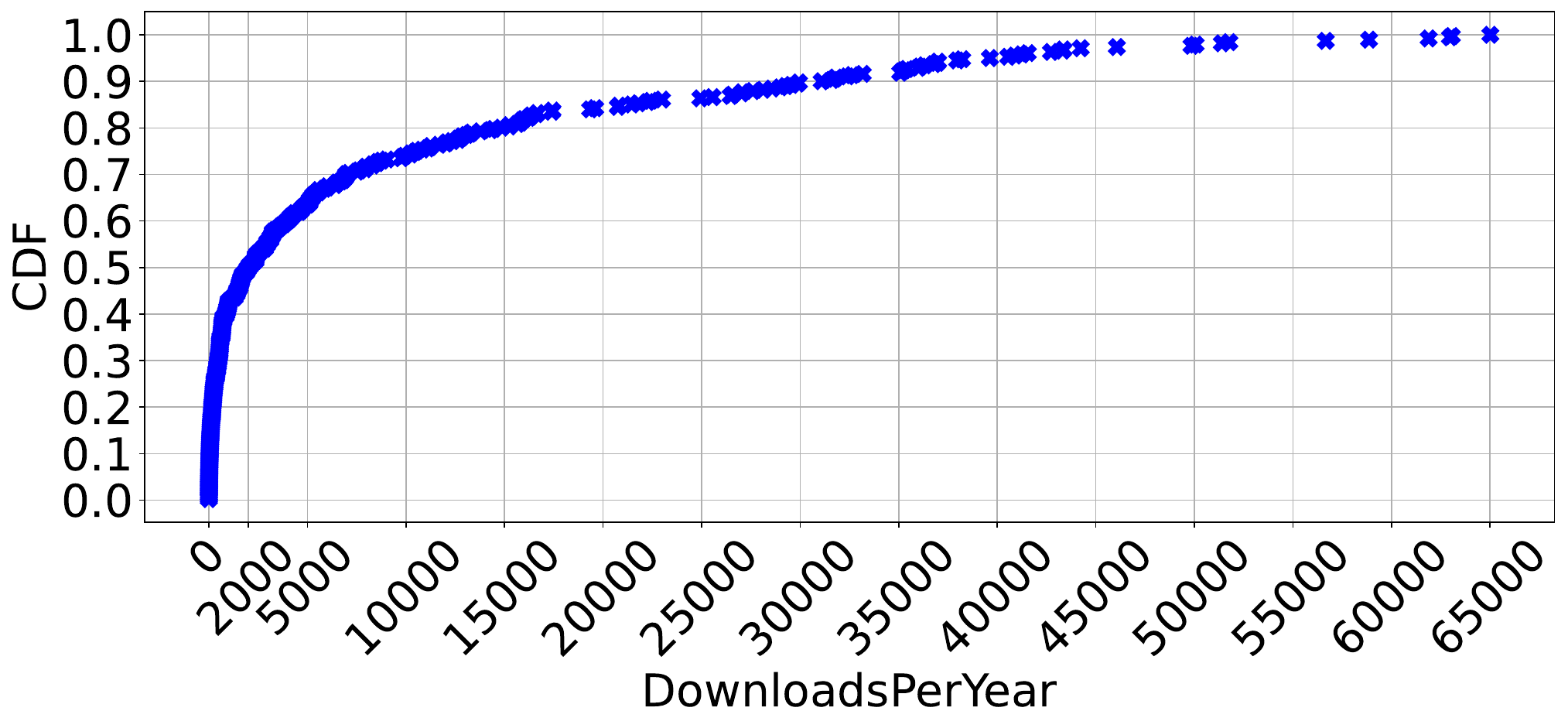}
    \caption{CDF of \texttt{DownloadsPerYear} after removing outliers.}
    \label{fig:cdf_dpy_no_outlier}
\end{figure}

We apply the Interquartile Range (IQR) method for outlier detection to remove the outliers from the dataset. This approach filters out 65 apps, leaving us with 381 apps for the analysis, with \texttt{DownloadsPerYear}. The CDF of \texttt{DownloadsPerYear} after the outlier removal is plotted in Figure \ref{fig:cdf_dpy_no_outlier}.

Next, we have three feature sets similar to the previous RQ. The \texttt{Voting} feature selection method identified nine key features: \texttt{maxNestedBlocksQty\_70\_class},   \texttt{finalFieldsQty\_95},\texttt{Storage}, \texttt{wmc\_40\_class}, \texttt{maxNestedBlocksQty\_70\_class} , \texttt{Photos/Media/Files},  \texttt{tryCatchQty\_max\_method},  \texttt{returnQty\_30}, \texttt{lcom\_90},  and \texttt{fanin\_20\_class} .

With these feature sets, we apply 6 regression models. As shown in Table~\ref{tab:dpy_regression}, all models exhibit high error values and low $R^2$ scores, indicating limited predictive power across the board. Among them, the Ridge regression model using the \texttt{Voting} feature set achieved the lowest RMSE (13,030.70) and the highest $R^2$ (0.08), offering the most consistent performance across all metrics. Although the overall results remain unsatisfactory, we select Ridge regression for further analysis due to its relatively stronger performance and greater stability compared to other models.

\begin{table}
\centering
\caption{Regression metrics (RMSE, MAE, $R^2$) of \texttt{DownloadsPerYear} regressors evaluated using three feature selection strategies: Size-only, Handpicked, and Voting. Bold values indicate the best metric scores within each feature set.}
\label{tab:dpy_regression}
\resizebox{\textwidth}{!}{
\begin{tabular}{lcccccccccc}
\toprule
\multirow{2}{*}{\textbf{Model}} & 
\multicolumn{3}{c}{\textbf{Size-only}} & 
\multicolumn{3}{c}{\textbf{Handpicked}} & 
\multicolumn{3}{c}{\textbf{Voting}} \\
\cmidrule(lr){2-4} \cmidrule(lr){5-7} \cmidrule(lr){8-10}
& \textbf{RMSE} & \textbf{MAE} & \textbf{$R^2$}
& \textbf{RMSE} & \textbf{MAE} & \textbf{$R^2$}
& \textbf{RMSE} & \textbf{MAE} & \textbf{$R^2$} \\
\midrule
MLP & 15549.39 & \textbf{8320.82} & -0.31 & 15292.60 & \textbf{8141.81} & -0.26 & 14887.51 & \textbf{7933.33} & -0.20 \\
Lasso & 13589.72 & 9876.01 & 0.00 &\textbf{ 13957.50} & 9839.24 & \textbf{-0.05} & 13031.64 & 8998.82 & 0.08 \\
Ridge & \textbf{13589.72} & 9876.01 & \textbf{0.00} & 13981.40 & 9851.91 & -0.06 & \textbf{13030.70} & 8998.74 & \textbf{0.08} \\
DT & 17628.71 & 11013.86 & -0.68 & 19261.85 & 12187.56 & -1.01 & 17924.23 & 10699.23 & -0.74 \\
RF & 15478.00 & 10241.84 & -0.29 & 14288.05 & 10545.20 & -0.10 & 13631.60 & 9594.51 & -0.00 \\
GB & 14600.80 & 9756.98 & -0.15 & 14846.89 & 10570.74 & -0.19 & 14330.94 & 9708.88 & -0.11 \\
\bottomrule
\end{tabular}
}
\end{table}

\begin{figure}[ht]
    \centering
    \includegraphics[width=0.75\linewidth]{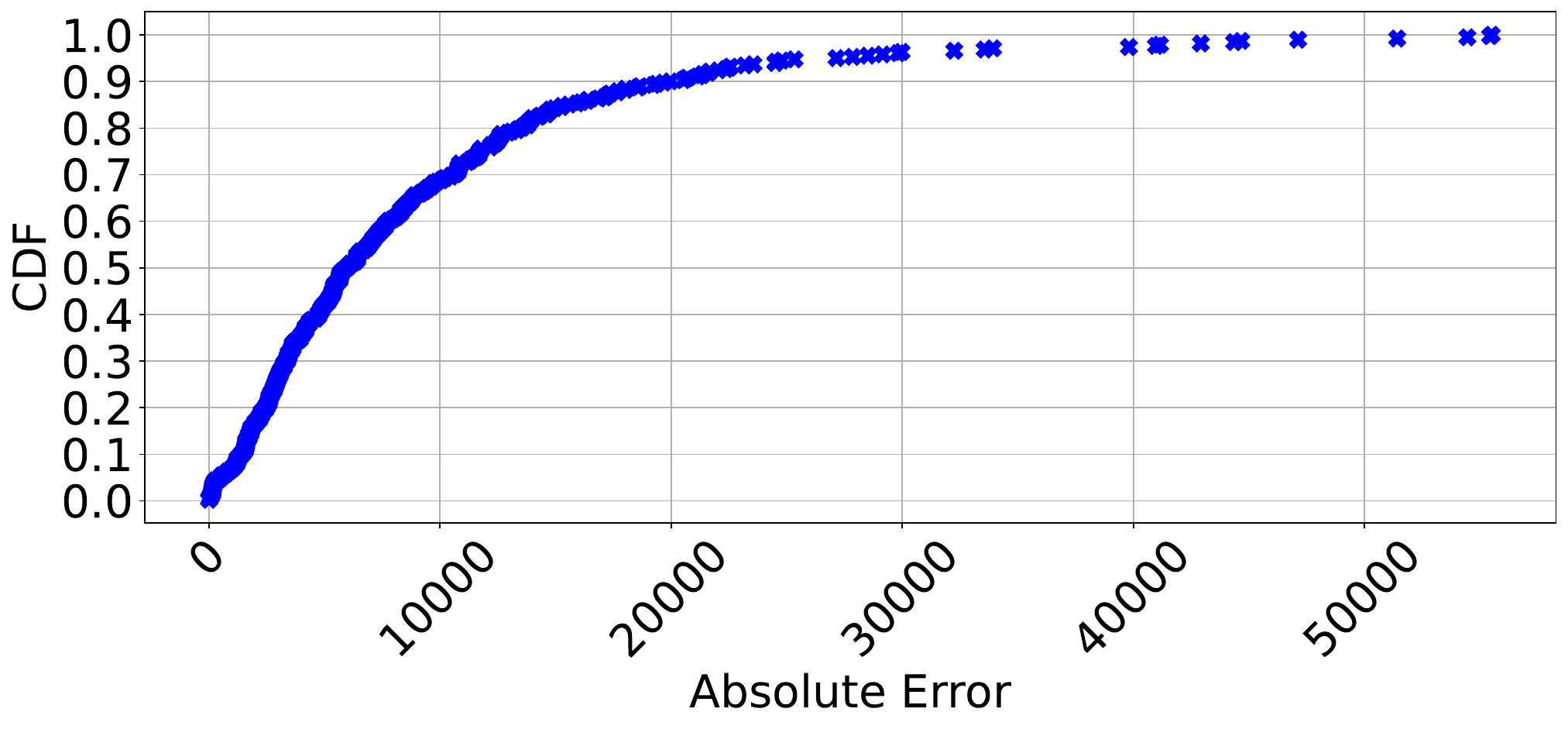}
    \caption{CDF of absolute error for the Ridge regression model in predicting \texttt{DownloadsPerYear}.}
    \label{fig:regression_dpy_abserror_ridge}
\end{figure}

The Cumulative Distribution Function (CDF) of Absolute Errors in Figure \ref{fig:regression_dpy_abserror_ridge} reveals that for 60\% of the samples, the absolute error exceeds 5000, which is significantly large in the context of \texttt{DownloadsPerYear}. The model achieves an absolute prediction error below 1,000 for only about 5\% of the samples, indicating that accurate predictions are rare and overall predictive accuracy is poor.

\begin{figure}[ht]
    \centering
    \includegraphics[width=0.6\linewidth]{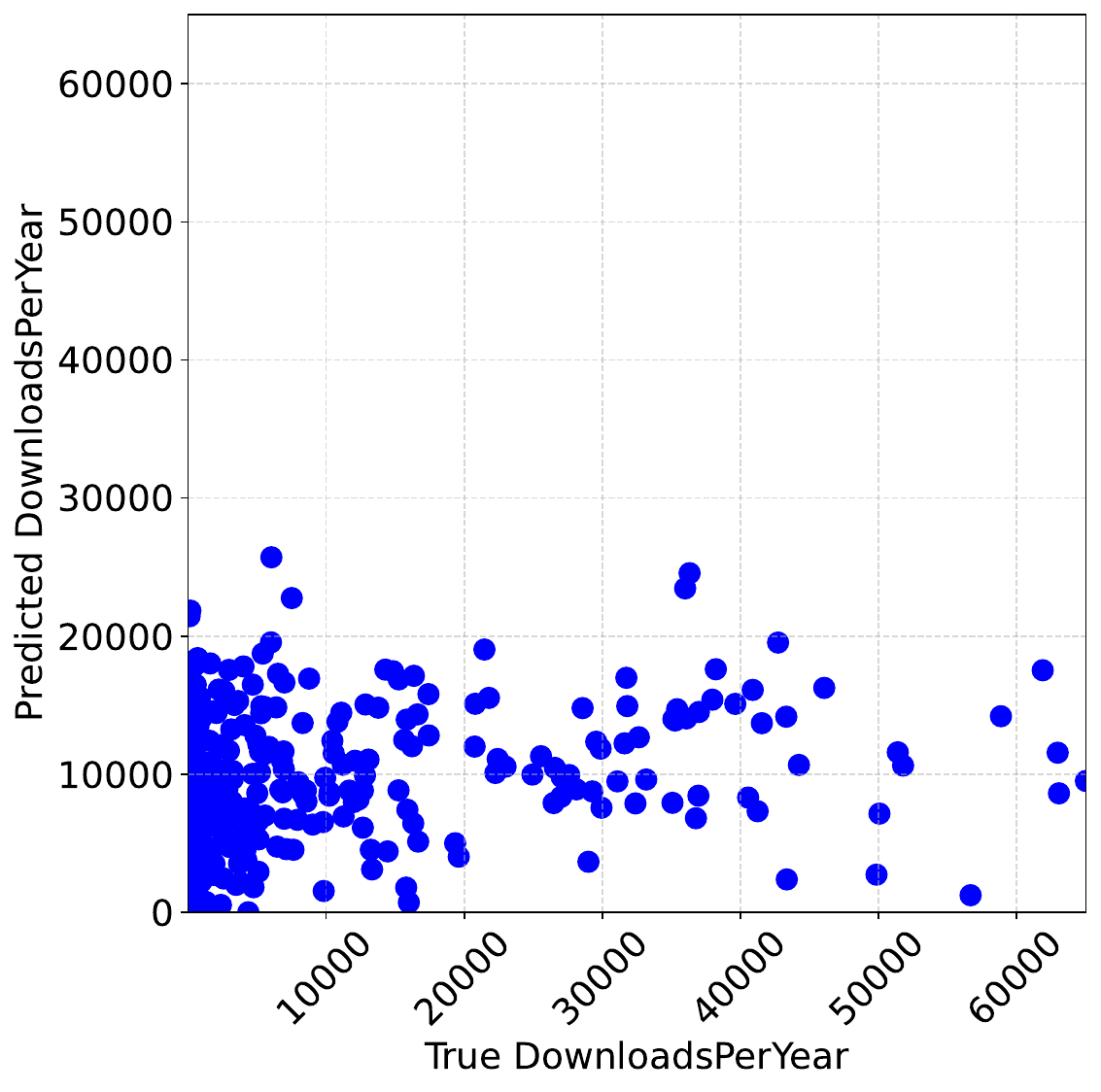}
    \caption{Scatter plot of true \texttt{DownloadsPerYear} vs predicted \texttt{DownloadsPerYear}.}
    \label{fig:regression_dpy_scatter}
\end{figure}

To further examine model performance, we analyze the scatter plot of true versus predicted values in Figure~\ref{fig:regression_dpy_scatter}. In a well-performing regression model, the data points should align closely along the diagonal, indicating accurate predictions. However, as shown in the figure, the predicted values are heavily concentrated below 20,000, despite the true values spanning a much wider range. This compression of output highlights the model's inability to capture the full variability in \texttt{DownloadsPerYear}. These findings reinforce the conclusion that regression models, when relying solely on internal software metrics, perform poorly.
 
In the previous research question, internal software metrics proved ineffective for regression due to the skewed distribution of app ratings. However, in this research question for \texttt{DownloadsPerYear}, the limitation stems from a different issue: the target variable spans a wide range—from 0 to over 65,000—across only 381 apps, making accurate value prediction challenging. Once again, internal metrics failed to provide reliable regression performance, prompting us to reframe the problem as a classification task.
\subsubsection{\textbf{Classification: Predicting Popularity Based on \texttt{DownloadsPerYear}.}}

We repeated the classification experiments using \texttt{DownloadsPerYear} as the popularity indicator. Based on the cumulative distribution function shown in Figure \ref{fig:cdf_dpy_no_outlier}, we defined a threshold of 2000 downloads per year, categorizing apps with \texttt{DownloadsPerYear} $\geq 2000$ as \emph{Popular} and those below 2000 as \emph{Unpopular}. The threshold of 2000 splits the dataset almost exactly in half, with 191 Popular and 190 Unpopular apps. 

The \texttt{Voting} method selected five features: \texttt{genreId\_TOOLS}, \texttt{loc\_min\_method}, \texttt{publicFieldsQty\_90}, \texttt{staticMethodsQty\_95}, and \texttt{Photos/Media/Files}. With three feature sets: \texttt{Size-only}, \texttt{Handpicked}, and \texttt{Voting}, we train and evaluate the classification models.
As the dataset is already balanced, resampling techniques are not required in this case. Table~\ref{tab:bin_dpy_imb_model_performance} presents the classification results, where the \texttt{MLPClassifier} achieved the best overall performance on the \texttt{Voting} feature set, with the highest MCC (0.37) and a strong AUC of 0.71. Although other models, such as Logistic Regression, attained slightly higher AUC values on specific feature sets, MLPClassifier demonstrated the most balanced performance across all three metrics within the Voting configuration.

\begin{table*}[htbp]
\centering
\caption{Classification metrics (Macro F1 Score, MCC, AUC) of \texttt{DownloadsPerYear} classifiers evaluated using three feature selection strategies: Size-only, Handpicked, and Voting. Bold values indicate the best metric scores within each feature set.}
\label{tab:bin_dpy_imb_model_performance}
\resizebox{\textwidth}{!}{
\begin{tabular}{lccccccccc}
\toprule
\multirow{2}{*}{\textbf{Model}} & 
\multicolumn{3}{c}{\textbf{Size-only}} & 
\multicolumn{3}{c}{\textbf{Handpicked}} & 
\multicolumn{3}{c}{\textbf{Voting}} \\
\cmidrule(lr){2-4} \cmidrule(lr){5-7} \cmidrule(lr){8-10}
& \textbf{Macro F1} & \textbf{MCC} & \textbf{AUC}
& \textbf{Macro F1} & \textbf{MCC} & \textbf{AUC}
& \textbf{Macro F1} & \textbf{MCC} & \textbf{AUC} \\
\midrule
MLP & 0.54 & 0.13 & 0.54 & 0.57 & 0.14 & 0.57 & \textbf{0.69} & \textbf{0.37} & 0.71 \\
LR & 0.49 & 0.09 & 0.47 & 0.55 & 0.10 & 0.59 & 0.67 & 0.33 & \textbf{0.72} \\
DT & 0.57 & 0.14 & 0.57 & 0.55 & 0.11 & 0.55 & 0.58 & 0.16 & 0.58 \\
RF & 0.57 & \textbf{0.15} & \textbf{0.57} & \textbf{0.59} & \textbf{0.19} & \textbf{0.64} & 0.62 & 0.25 & 0.65 \\
GB & 0.56 & 0.11 & \textbf{0.58} & \textbf{0.59} & 0.18 & 0.60 & 0.66 & 0.32 & 0.70 \\
\bottomrule
\end{tabular}
}
\end{table*}

\begin{figure}[ht]
    \centering
    \includegraphics[width=\textwidth]{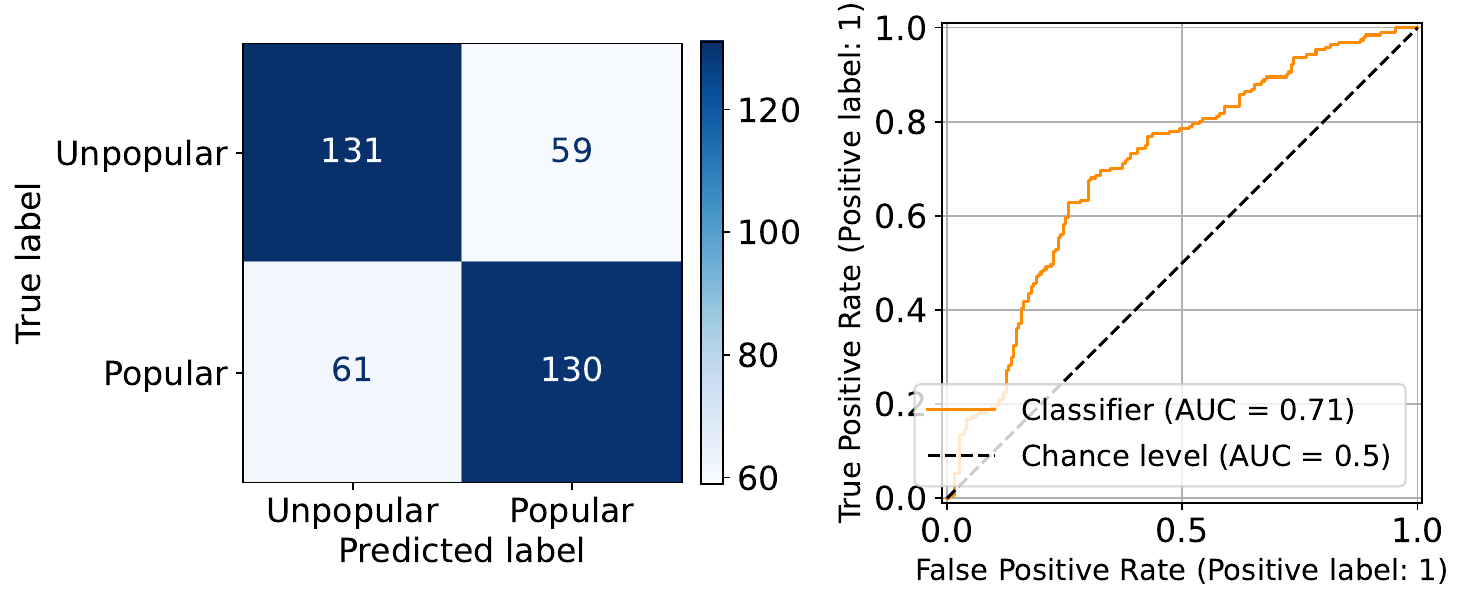}
    \caption{Confusion matrix and AUC-ROC curve for \texttt{DownloadsPerYear}-based popularity prediction using MLP classifier.}
    \label{fig:dpy_mlp}
\end{figure}

The class-wise performance metrics and confusion matrix are presented in Table \ref{tab:classification_report_dpy_mlp} and Figure \ref{fig:dpy_mlp}, respectively. The model achieves a precision of 0.682 and a recall of 0.689 for the Unpopular category, leading to an F1-score of 0.685. Meanwhile, for the Popular category, the model attains a precision of 0.687 and a recall of 0.680, resulting in an F1-score of 0.684. These results indicate that the model maintains a balanced performance across both classes, unlike what we observed in RQ1 for popularity based on ratings. Overall, these results again confirm the usefulness of internal metrics.  

\begin{table}[ht]
\centering
\caption{Classification performance for \texttt{DownloadsPerYear}-based popularity prediction using MLP classifier.}
\begin{tabular}{lccc}
\hline
\textbf{Class}      & \textbf{Precision} & \textbf{Recall}   & \textbf{F1-Score} \\ \hline
Unpopular     & 0.682     & 0.689    & 0.685    \\
Popular       & 0.687     & 0.680    & 0.684    \\ \hline
\end{tabular}
\label{tab:classification_report_dpy_mlp}
\end{table}

\begin{summarybox}{Summary of RQ2}
    Regression models using internal software metrics struggled to predict \texttt{DownloadsPerYear} accurately, with high error rates and low $R^2$ values, even after outlier removal. Reframing the task as a classification problem improved performance modestly, with the MLP classifier achieving balanced precision and recall, suggesting some potential for distinguishing between popular and unpopular apps.
\end{summarybox}

\section{Discussion}
\label{sec:discussion}


Our empirical investigation into the predictive power of internal software metrics for application popularity reveals both promising insights and notable limitations. In addressing RQ1—predicting application ratings via regression—we found that the models exhibited limited generalizability, reflected in low $R^2$ scores, and failed to capture the underlying variability in the data. However, when reframed as a classification task (i.e., distinguishing popular from unpopular apps), model performance improved.  These results suggest that internal software metrics carry meaningful predictive value at release time. A similar pattern emerged in RQ2, which examined the prediction of the \emph{DownloadsPerYear} metric.

It is unsurprising that software metrics alone do not fully explain application popularity. Clearly, non-code factors, such as developer reputation, historical performance, the level of competition for the app, the usability of its UI, and marketing efforts, will likely play a significant role in shaping popularity outcomes~\citep{tian2015characteristics}. Incorporating these features in future work may lead to more accurate and comprehensive predictive models. Considering the potential influence of the non-code factors, it is striking that internal software metrics can still offer valuable insight into app popularity. What seems obvious from this study is that a development organization would be unwise to ignore software metrics in deciding whether a product is ready for release.




\subsubsection*{\textbf{Implications for Practitioners}}
Our findings suggest that different internal metrics can contribute signals for predicting app popularity. Practitioners should therefore view code-level metrics not only as maintainability indicators but also as potential predictors of market performance. We recommend that software development teams systematically incorporate code quality monitoring into their development processes. The integration of static analysis tools capable of tracking complexity, coupling, and code smells can aid in identifying latent issues that may impact both maintainability and user perception. Early detection of such issues allows teams to implement preventive measures, thereby improving both software quality and user satisfaction and potentially warding off technical debt \citep{cai2023software}. Furthermore, practitioners are encouraged to explore the use of machine learning and predictive analytics as part of their quality assurance strategy.

\subsubsection*{\textbf{Implications for Researchers}}
From a research standpoint, our findings provide meaningful insights into the role of internal software metrics in predicting application popularity. Contrary to earlier studies~\citep{el_emam_confounding_2001,herraiz_towards_2007,gil_correlation_2017,scalabrino_improving_2016} that positioned code size as the principal---or sole---predictor of software quality and performance, our findings reveal that a wider range of metrics also contribute meaningfully to predictive accuracy, especially when context is considered as the king (regression vs. classification), as was also concluded by some other relevant studies---class-level vs method-level evaluation~\citep{landman_empirical_2014,chowdhury2025evidence}, and \#revisions vs. change size based evaluation ~\citep{chowdhury_revisiting_2022}. 

These findings should encourage the research community to develop a more comprehensive set of code metrics and to evaluate them based on different contexts. Additionally, the improved performance achieved using feature selection algorithms underscores the importance of this preprocessing step in predictive modeling. These findings should motivate future work on designing and evaluating more sophisticated feature selection algorithms to further enhance prediction performance. We also observed that a model's performance is significantly impacted by the type of algorithm used for addressing the class imbalance problem, indicating a strong need for additional research in this direction. Researchers may also focus on building genre-specific models rather than generic ones, as genre type emerged as an important feature in both of our classification models. However, the small number of apps that are commonly available on both F-Droid and the Google Play Store would be a challenge for building such models.   


\subsection{Threats to Validity}

The \emph{external validity} of our study is limited by our focus on Android apps exclusively. As a result, the findings may not be generalizable to iOS apps. Additionally, within the Android ecosystem, we concentrated solely on Java-based applications, although other programming languages, such as Kotlin, are also commonly used to develop Android apps. We can not imagine any reason why the difference in platform or language would significantly affect our findings, but it still remains an external threat.

Also, our dataset is relatively small, as we specifically needed apps available on both F‑Droid and the Google Play Store. However, linking these two repositories has been the most viable approach when access to both source code and popularity indicators were required~\citep{grano_android_2017,catolino_does_2018,coppola2019characterizing}. 

\textit{Internal validity} may be affected by the presence of confounding variables, such as user reviews, marketing efforts, pricing strategies, and external events like app store promotions, which were not fully controlled in our analysis. Although we tried to collect the oldest available version for each app with the goal of predicting popularity at the initial stage of development, the oldest version available on F-droid may not be the first or the initial version. 

\textit{Construct validity} also poses a challenge, as the operational definitions of “code quality” and “app popularity” may not fully capture their conceptual meanings. While app popularity was measured primarily through downloads and ratings, factors such as user engagement, revenue, and retention rates could offer a more holistic perspective. Similarly, relying on a specific method, such as static code analysis, to measure code quality introduces the risk of mono-method bias, as alternative tools or frameworks might yield different insights.

\textit{Conclusion validity} of our study is influenced by the various threats outlined above. Although the inclusion of genre-level features provides valuable insights, the findings may not fully account for variations within individual app genres, underscoring the need for more granular analysis to better understand genre-specific patterns.

\section{Conclusion}
\label{sec:conclusion}
This study investigates the relationship between internal software metrics and app popularity indicators, specifically user ratings and download counts. While our results indicate that internal software metrics alone are insufficient for accurately predicting continuous measures such as ratings or yearly downloads, they demonstrate predictive power when app popularity is reformulated as a binary classification problem. These findings highlight the potential utility of internal code metrics in understanding and characterizing software success.

Our study underscores the potential of techniques, such as predictive modeling, to augment the evaluation and forecasting of app success. These methodologies offer promising directions for identifying critical determinants of performance and fostering innovation in software development practices. For practitioners, our findings advocate for sustainable and quality-driven coding practices aligned with market demands. For researchers, the results affirm the relevance of internal software metrics and encourage further exploration into their role, beyond size, in shaping software outcomes.

In summary, this work contributes to the growing body of research at the intersection of software engineering and market-oriented evaluation, advancing our understanding of how technical quality metrics relate to app success in competitive marketplaces.


\bibliographystyle{spbasic}
\bibliography{sample-base, references}

\end{document}